\documentclass[prb,superscriptaddress,twocolumn, floatfix]{revtex4-2}
\usepackage{booktabs}
\usepackage{makecell}
\usepackage{mathtools}
\usepackage{verbatim}
\usepackage{graphicx}
\usepackage{xcolor}
\usepackage{amsmath}
\usepackage{amsfonts}
\usepackage{amsthm}
\usepackage{amssymb}
\usepackage{pifont}
\usepackage{amsbsy}
\usepackage{wasysym}
\usepackage{bm}
\usepackage{mathrsfs}
\usepackage{color}
\usepackage{hyperref}
\usepackage{soul}
\usepackage{multirow}
\usepackage{bbm}
\usepackage{physics}
\hypersetup{
    colorlinks=true,
    linkcolor=red,        
    citecolor=blue,
    breaklinks=true,
    urlcolor=blue
}
\DeclareMathAlphabet{\mathcal}{OMS}{zplm}{m}{n}

\newcommand{\sn}{\mathrm{sn}}
\newcommand{\cn}{\mathrm{cn}}
\newcommand{\dn}{\mathrm{dn}}
\newcommand\scalemath[2]{\scalebox{#1}{\mbox{\ensuremath{\displaystyle #2}}}}

\newcommand{\opS}[2]{\hat{S}_{#2}^{#1}}
\newcommand{\ops}[2]{\hat{s}_{#2}^{#1}}
\newcommand{\cmark}{\ding{52}}%
\newcommand{\xmark}{\ding{56}}%

\usepackage{orcidlink}

\begin{document}

\title{Granovskii-Zhedanov Scars of XYZ Models: Modern Algebraic Perspectives and Realization in Higher Dimensional Lattices}

\author{Dhiman Bhowmick}
\altaffiliation{\href{mailto:dhiman002@e.ntu.edu.sg}{dhiman002@e.ntu.edu.sg}}
\affiliation{Department of Physics, National University of Singapore, Singapore 117551}

\author{Wen Wei Ho}
\altaffiliation{\href{mailto:wenweiho@nus.edu.sg}{wenweiho@nus.edu.sg}}
\affiliation{Department of Physics, National University of Singapore, Singapore 117551}
\affiliation{Centre for Quantum Technologies, National University of Singapore, 3 Science Drive 2, Singapore 117543}



\date{\today}

\begin{abstract}
In a work by Granovskii and Zhedanov, a surprising family of scar states exhibiting zero entanglement was discovered in the XYZ spin chain—remarkably, nearly three decades before the concept of many-body scars became a subject of active research.
Despite its significance, these states have largely gone unnoticed within the physics community.
In this study, we uncover the origin of the family of Granovskii-Zhedanov (GZ) scars within the framework of the modern algebraic understanding of quantum many-body scars.
We demonstrate that the scar subspace can be effectively described using the spectrum-generating algebra (SGA) framework, as well as through a group-theoretical formulation of the XXZ Hamiltonian. 
This description, however, is strictly applicable only in the XXZ limit, where a quasi-$U(1)$ symmetry exists within the scar subspace.
In contrast, the absence of such quasi-$U(1)$ symmetry in the GZ scar subspace restricts the direct applicability of these standard formulations. 
To address this, we adopt three alternative approaches.
First, we perturbatively extrapolate an approximate SGA for the XYZ system from the XXZ system.
Second, we construct the standard SGA directly from the GZ states in the XYZ limit.
In the third approach, we numerically optimize the SGA generator and demonstrate that, apart from special $q$-values, the optimized generator is a local operator with support on two nearest-neighbor sites.
Employing these algebraic constructions, we identify the scar subspaces of the XXZ and XYZ systems and clarify their interrelationships.
We further explore the possibility of constructing lattice-independent GZ scars in higher-dimensional uniform spin-exchange systems with centrosymmetry, using graphical rules developed for GZ scar construction.
Our results indicate that lattice-independent GZ scars cannot be supported on uniform lattices with odd coordination numbers or plaquettes with an odd number of edges, while uniform lattices featuring even coordination numbers and even-edged plaquettes can host such lattice-independent scars in specific scenarios.
Remarkably, if certain bonds retain the full $SU(2)$ symmetry of the spin-exchange interaction—thereby breaking the spatial uniformity of the lattice—lattice-independent GZ scars can still emerge in systems with odd coordination numbers or plaquettes with an odd number of edges.
\end{abstract}

\maketitle


\section{Introduction}

The Eigenstate Thermalization Hypothesis (ETH) is a cornerstone in quantum many-body dynamics\,\cite{ETH1,ETH2,ETH3,ETH4}.
ETH has long been thought to describe the properties of all excited states in the middle of the spectrum of any generic quantum many-body system\,\cite{ETH_more1, ETH_more2, ETH_more3,ETH_more4,ETH_more5}.
Although the formulation of ETH appears to be generic, it is nevertheless a hypothesis, and thus, it is natural to ask whether potential counterexamples exist.
Many-body localization\,\cite{MBL1,MBL2} and integrable systems strongly violate the ETH, whereas quantum many-body scars (QMBS) constitute a weak violation, as only an exponentially small subset of states in the Hilbert space deviate from it.

The first many-body scar is observed in a Rydberg atom quantum simulator, where the simplistic product states $\mathbb{Z}_2$ revealed unexpected long-lived revivals of the state\,\cite{PXP1,PXP2}. 
These revivals were linked to a small subset of non-thermal eigenstates QMBS of the non-integrable PXP model used to describe the system.
As a result, the local quantum information stored in the system remains intact without dissipating through thermalization, making it a promising candidate for applications in quantum information and quantum computing\,\cite{QuantumSensing, DecoherenceFreeScar}.
Subsequently, numerous studies emerged, aiming to understand the origin of QMBS and explore ways to stabilize them within the paradigmatic PXP model\,\cite{PXP_more1, PXP_more2, PXP_more3, PXP_more4, PXP_more5, Stabilize1, Stabilize2}.
The current knowledge of this phenomena is not only limited to the PXP chain, variety of other interesting systems have also been identified, including the 1D transverse field Ising model\,\cite{Ising1,Ising2}, the fermionic Hubbard model\,\cite{Hubbard1, Hubbard2, Hubbard3, Hubbard4}, quantum Hall systems\,\cite{Hall1, Hall2}, fracton topological ordered systems\,\cite{Fracton1, Fracton2}, AKLT spin chain\,\cite{AKLT1, AKLT2, AKLT3, AKLT4}, the spin-1 XY model\,\cite{XY1, XY2}, frustrated spin systems\,\cite{FrustratedSpin1, FrustratedSpin2} and more\,\cite{Review1, Review2}.

Spin systems, in particular, have long been a central focus in the study of exotic phenomena in many-body physics, owing to their straightforward implementation in both condensed matter systems\,\cite{Kitaev1, Kitaev2, Kitaev3, Haldane1, Haldane2, SSModel1, SSModel2} and quantum simulators\,\cite{Simulator1, Simulator2, XXZ_chain1, XXZ_chain2, XXZ_chain3}.
Interestingly, the spin-$S$ XXZ chain supports a family of scar states, referred to as the helical scars, which constitute zero-entanglement periodic product eigenstates.
Experimentally, these states have also been realized in cold atom systems for the spin-$1/2$ XXZ chain, where it has been identified as a Bethe-phantom state\,\cite{PhontomBethe_PeriodicBoundary,PhontomBethe_OpenBoundary, XXZ_chain2, XXZ_chain3}.
The helical states have been a central focus of numerous studies due to its anomalous dynamical behavior\,\cite{HelixDynamics1,HelixDynamics2, HelixDynamics3, HelixDynamics4, GeneralizedSpinHelix, FelixGraphicalConstruction}.
Surprisingly, an analogous family of states persists in the spin-$S$ XYZ chain, even in the absence of a $U(1)$ symmetry.
Originally introduced by Granovskii and Zhedanov, these scar states are hereafter referred to as the Granovskii–Zhedanov (GZ) scars\,\cite{GranovskiiZhedanov1, GranovskiiZhedanov2}.
Although the work of Granovskii and Zhedanov was contemporaneous with Heller’s introduction of quantum scars in the single-particle context\,\cite{Heller}, the concept of many-body scars did not emerge in the literature until nearly three decades later.
This naturally prompts the question of how scars of such striking character, yet long overlooked by the physics community, fit into the contemporary understanding of many-body scar phenomena.
These states have only recently been reintroduced in our work\,\cite{AssyemtricDecay}, as well as independently in a study by Mingchen et al.\,\cite{ExactSpinHelix}.
However, many fundamental questions about the nature and structure of the GZ states remain unresolved, which we aim to address in the present study.

In this study, we investigate the scar subspace associated with the GZ scars. 
We demonstrate that this subspace possesses a thermodynamically large degeneracy, with the number of degenerate states scaling linearly with both the system size $N$ and the spin value $S$.
Specifically, we show that the dimension of the scar subspace is exactly $4NS$.
It is well established in the literature that the scar subspace underlying many-body scar states can be understood within several theoretical frameworks, including Spectrum Generating Algebra (SGA), Krylov-restricted thermalization, projector embedding, and, more recently, Fock-space caging\,\cite{Review1, SGA2, AKLT3, Hubbard4, GroupTheory1, GroupTheory2, FCS1, FCS2, FCS3}.

We show that, in the XXZ limit, the scar subspace admits a description in terms of a conventional SGA.
In this limit, the GZ scars reduce to the more widely studied helical scars\,\cite{HelixDynamics1,HelixDynamics2, HelixDynamics3, HelixDynamics4, GeneralizedSpinHelix, FelixGraphicalConstruction,XXZ_chain1, XXZ_chain2, XXZ_chain3, XXZ_chain4}. 
Furthermore, the $4NS$-dimensional scar subspace decomposes into two positive- and negative-helicity helical scar subspaces, each of dimension $2NS+1$. 
These two sectors are independently described by separate SGAs corresponding to opposite helicities.
We also demonstrate that the resulting helical scar subspace is equivalent to a set of bosonic $\zeta$-states and the Hamiltonian breaks down into $\hat{H}_0 + \sum_a \hat{O}_a \hat{T}_a$ group theoretical structure explaining the existence of such scar subspace\,\cite{GroupTheory1, GroupTheory2}.

However, in the XYZ limit, the absence of $U(1)$ symmetry prevents the application of the standard SGA formalism associated with the XXZ case.
To overcome this limitation, we employ three alternative approaches, each possessing distinct advantages and drawbacks.
First, we perturbatively extrapolate the SGA formalism for the XYZ case from the XXZ system, and this procedure enables an approximate analytical SGA construction.
Second, we construct the exact SGA directly starting from the GZ states, which, although being exact, lacks the analyticity as well as the locality of the SGA generator.
In the third and final approach, we numerically optimize an SGA ansatz based on the SGA commutation relation.
Interestingly, in this last approach, we find that the $4NS$-dimensional GZ scar subspace, except at two special parameter points, can be generated by a local generator made of spin operators with support extending only over at most two sites.

These approaches not only enable the characterization of the SGA but also uncover rich internal structures and nontrivial connections between the GZ scar subspaces in the XXZ and XYZ limits.
Notably, we find that the span of the XYZ GZ scar subspace with a particular helicity is larger than that of the helical scar subspace with the same helicity in the XXZ limit. 
This observation indicates that the XYZ GZ scars with a particular helicity cannot be understood merely as deformations of the helical scars belonging to the corresponding helicity sector.
Consequently, the standard SGA formalism valid for the XXZ helical scar subspace cannot be straightforwardly generalized to the more general GZ scar subspace in the XYZ model.
Nevertheless, the GZ states with a fixed helicity in the XYZ limit retain a substantial overlap with the corresponding helical scar subspace of the XXZ limit, suggesting that the GZ scar states admit an approximate representation in the XXZ helical scar basis.
The overlap between the GZ states and the helical scar subspace is strongest near $\gamma = 0$ and becomes significantly reduced as $\gamma$ approaches $1$, where $\gamma$ denotes a free parameter in the definition of the GZ states introduced later.

We further demonstrate that the presence of the GZ scars is not limited to a one-dimensional spin-$S$ XYZ spin chain, but can also arise in higher-dimensional lattice systems.
Felix et al.\,\cite{FelixGraphicalConstruction} show that the helical scars of XXZ models can be constructed in arbitrary-dimensional lattices using simple graphical rules: the vertex rule and the circuit rule.
These laws are also applicable within the XYZ limit, with slight modifications.
Applying these rules, we identify the classes of uniform lattices in two dimensions with centrosymmetric spin-exchange (CSSE) interactions that support the construction of lattice-independent GZ scars, where lattice-independent means that the choice of the parameters of the scar states is independent of the lattice geometry and solely depends on the system Hamiltonian.
We also demonstrate that uniform lattices, which are unsuitable for constructing lattice-independent GZ scars, can still support such states, provided that certain bonds retain the full $SU(2)$-symmetric Heisenberg exchange interaction, thereby breaking the spatial uniformity of the system.

\begin{figure}[t]
\includegraphics[width=0.5\textwidth]{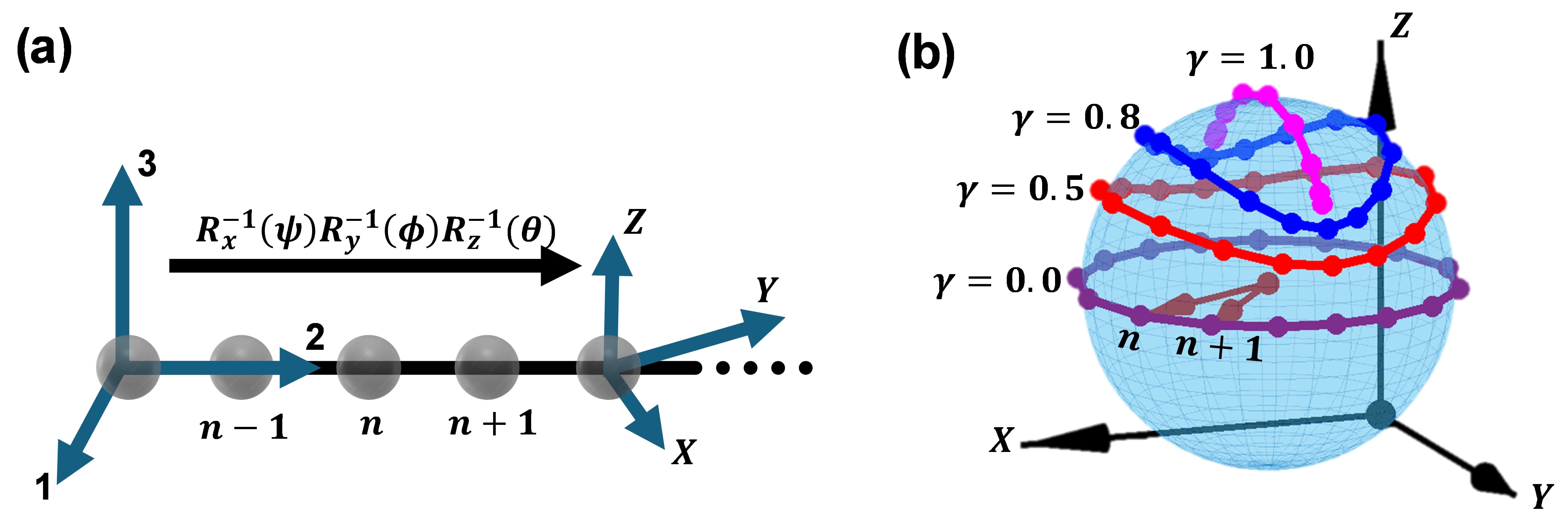}
\caption{(a) Schematic representation showing that the transformation from a generic NN CSSE Hamiltonian Eq.\,\ref{Eq::GenericCentroSymmetricHamiltonian} to an XYZ Hamiltonian \ref{Eq::XYZ_Hamiltonian} is equivalent to an $SO(3)$ rotation, where the rotation operators $(R_x(\psi), R_y(\phi), R_z(\theta))$ are acting on the lab frame $(1,2,3)$.
(b) The Bloch sphere of radius $S$ represents the Granovskii-Zhedanov scars on a 1D spin chain. The dots on the sphere denote spin sites, while the arrows extending from the origin to each dot represent the local spin states. 
}
\label{fig::1DChain}
\end{figure}

\section{Overview of GZ scars}
\label{Sec::Overview_of_GZ_scars}
A centro-symmetric model is one in which inversion symmetry is preserved.
A generic centro-symmetric spin exchange (CSSE) Hamiltonian with nearest-neighbor (NN) coupling on a 1D chain can be written as,
\begin{align}
    \hat{H}&=\sum_n 
    \begin{pmatrix}
        \opS{1}{n} \\
        \opS{2}{n} \\
        \opS{3}{n}
    \end{pmatrix}^T
    \begin{pmatrix}
        J_{11} & J_{12} & J_{13}\\
        J_{21} & J_{22} & J_{23} \\
        J_{31} & J_{32} & J_{33}
    \end{pmatrix}
    \begin{pmatrix}
        \opS{1}{n+1} \\
        \opS{2}{n+1} \\
        \opS{3}{n+1}
    \end{pmatrix},
    \label{Eq::GenericCentroSymmetricHamiltonian}
\end{align}
where $(1,2,3)$ represent the orthogonal directions in the lab frame\,(see Fig.\,\ref{fig::1DChain}(a)) and the operator $\opS{r}{n}$ denotes $r$-th component of the spin-$S$ operator at site $n$.
This centro-symmetric condition is satisfied by $J_{ij}=J_{ji}$.
This Hamiltonian can be transformed into the simple 1D XYZ Hamiltonian via diagonalization using $SO(3)$ matrices, which is equivalent to a change of reference frame (see Fig.\,\ref{fig::1DChain}(a)),
\begin{equation}
    \hat{H}=\sum_n \left[
    J_x \opS{x}{n}\opS{x}{n+1}
    +J_y \opS{y}{n}\opS{y}{n+1}
    +J_z \opS{z}{n}\opS{z}{n+1}
    \right].
    \label{Eq::XYZ_Hamiltonian}
\end{equation}
Granovskii and Zhedanov show that a family of periodic product states is a set of quantum many-body scars of the 1D XYZ spin chain (see Appendix.\,\ref{Appendix::Sec::ProofGZScar}),
\begin{align}
\label{Eq::GZScar}
    \ket{\Psi_{\pm GZ}}=
    \hat{\mathcal{R}}_{\pm}
     \ket{\Uparrow} \;,
\end{align}
where the sign in the subscript $+$ and $-$ denote the positive and negative helicity of a GZ scar state, respectively.
The rotation operators $\hat{\mathcal{R}}_\pm$ are defined as,
\begin{align}
\label{Eq::RotationOperator}
    &\hat{\mathcal{R}}_\pm(\gamma,\phi,q,\kappa)=\prod_n \exp(\pm i \opS{z}{n} \phi_n) \exp(-i \opS{y}{n} \theta_n)
    \nonumber \\
    &\text{with, }
    \theta_n=\cos^{-1}\left(\frac{\langle{\opS{z}{n}}\rangle}{S}\right), \;\;
    \phi_n=\text{atan}2\left(\langle{\opS{y}{n}}\rangle,\langle{\opS{x}{n}}\rangle\right),
\nonumber \\
    &
    \langle \opS{x}{n}\rangle=S\alpha\,\cn(nq+\phi,\kappa),\,
    \langle \opS{y}{n}\rangle=S\beta\,\sn(nq+\phi,\kappa),\,
    \nonumber\\
    &\;\;\;\;\;\;\;\;\;\;\;\;\;\;\;\;\;
    \langle \opS{z}{n}\rangle=S\gamma\,\dn(nq+\phi,\kappa),
\end{align}
where $\ket{\Uparrow}$ is a fully-polarized state and $(\cn, \sn, \dn)$ are Jacobi elliptic functions. The parameters $(q,\kappa)$ are related to the parameters of 1D XYZ Hamiltonian $(J_x, J_y, J_z)$ as follows (in the case $J_y\geq J_x >J_z$),
\begin{align}
    \dn(q,\kappa)=\frac{J_x}{J_y},\, \cn(q,\kappa)=\frac{J_z}{J_y},\,
    \kappa^2=\frac{J_y^2-J_x^2}{J_y^2-J_z^2}.
    \label{Eq::q_kappa_definition}
\end{align}
On the other hand, parameters $(\alpha, \beta, \gamma)$ are related as,
\begin{equation}
    \alpha=\sqrt{1-\gamma^2},\,
    \beta=\sqrt{1-\gamma^2+\kappa^2\gamma^2},\,
    |\gamma|\leq 1.
    \label{Eq::GammaDefinition}
\end{equation}
It is important to note that the parameters $\gamma$ and $\phi$ are independent of the spin system and can therefore be chosen arbitrarily. Consequently, Eq.\,\ref{Eq::GZScar} does not only represent a unique scar state but rather a family of scar states which are visualized in Fig.\,\ref{fig::1DChain}(b). 
Moreover, the scars represented by Eq.\,\ref{Eq::GZScar} are a generalization of the helical scars of easy-plane 1D XXZ spin chain to the XYZ model.
For a finite periodic chain of length $N$, the only allowed GZ scar states correspond to $q$-values of the form $4pK(\kappa)/N$, where $p$ is an integer and $K(\kappa)$ is the complete elliptic integral of the first kind.
Mingchen Zheng et al. recently rediscovered analogous states in spin-$S$ XYZ chains\,\cite{ExactSpinHelix}, where $\vartheta$ functions are used to represent the GZ states instead of the Jacobi elliptic functions employed in the present study. 
In Appendix\,\ref{Appendix::Sec::ThetaState}, we demonstrate how the conventions adopted in that work are related to those used in our study.

\begin{widetext}

\begin{figure}[t]
\centering
\includegraphics[width=\textwidth]{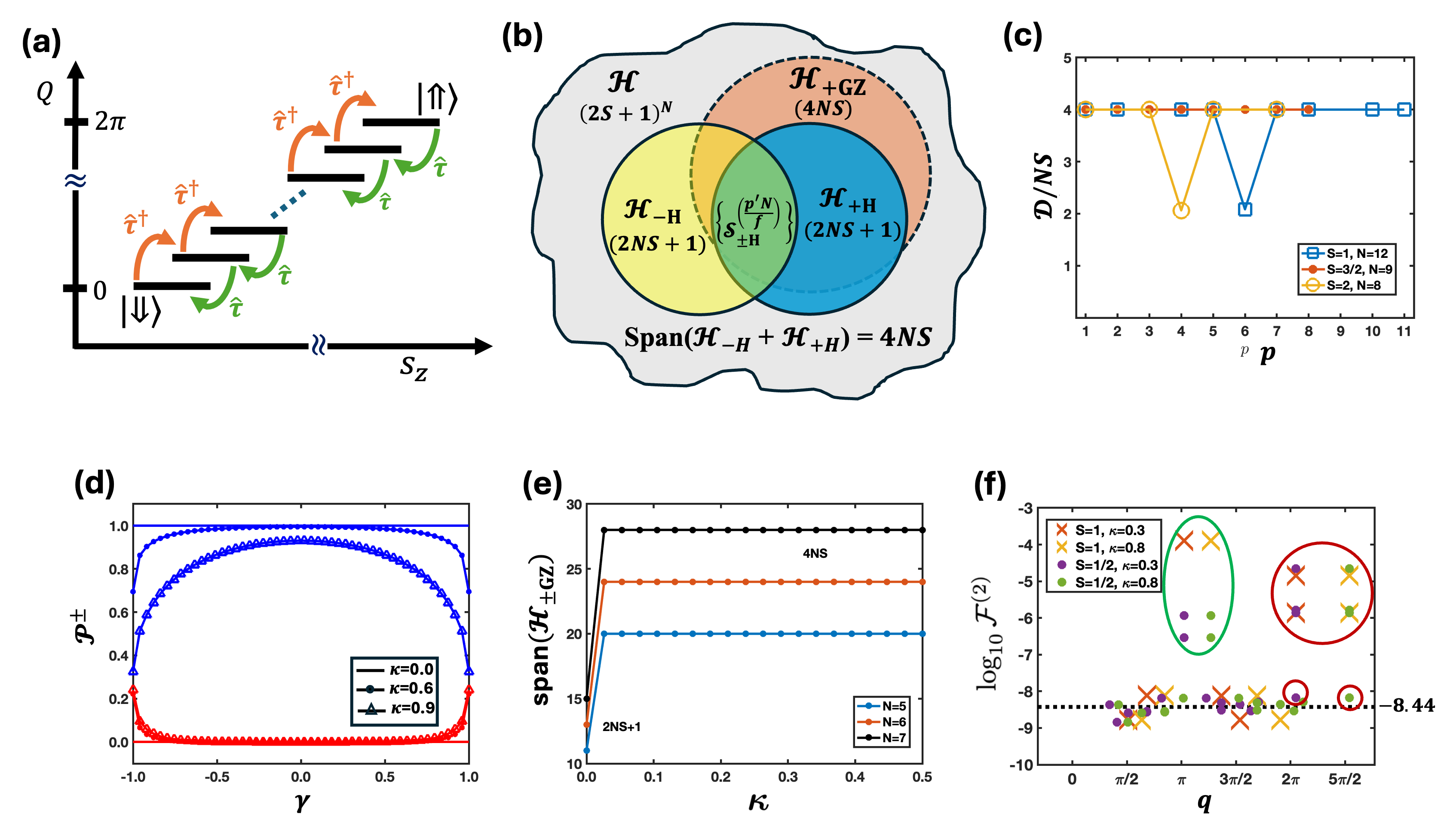}
\caption{
(a) The tower of helical scars in the XXZ spin chain is shown in the space of translational quantum number $Q$ and spin quantum number $s_z$.
The operator $\hat{\tau}$ ($\hat{\tau}^\dagger$) acts as a generator capable of producing all basis states within the helical scar subspace of a fixed helicity, starting from the fully polarized reference state $\ket{\Uparrow}$ ($\ket{\Downarrow}$).
(b) The schematic illustrates the structure and organization of the helical ($\mathcal{H}_{\pm \text{H}}$) and GZ scar ($\mathcal{H}_{\pm \text{GZ}}$) subspaces within the Hilbert space $\mathcal{H}$.
It is noteworthy that the schematics for the GZ scar subspaces with positive helicity, $\mathcal{H}_{+\text{GZ}}$, represent non-orthonormal states generated by the rotation operator (Eq.\,\ref{Eq::GZScar}), whereas the schematics for the helical scar subspaces with positive (negative) helicity, $\mathcal{H}_{+\text{H}}$ ($\mathcal{H}_{-\text{H}}$), represent orthonormal states generated by the SGA generator (Eq.\,\ref{Eq::XXZ_Scar_Subspace}).
(c) The degeneracy at the energy of GZ scars as a function of $q$-value ($q=4pK(\kappa)/N$) for various values of spins and system sizes.
The degeneracy is generally $4NS$, except in the case of spin $S=1/2$, or when $q$ takes special values that are integer multiples of $K(\kappa)$.
For these special $q$-values, the Hamiltonian turns into either the XY Hamiltonian or the isotropic Heisenberg Hamiltonian with either ferromagnetic or antiferromagnetic Ising interactions.
In particular, the value $\kappa=0.8$ is used for the numerical simulations. 
Data points corresponding to $q$-values that are odd integer multiples of $K(\kappa)$ are omitted from the plot for clarity, due to their high degeneracy (approximately 3000).
(d) Projection of the GZ scar state with positive helicity onto the spaces of helical scar states—with positive helicity ($\mathcal{P}^+$, shown in blue) and negative helicity ($\mathcal{P}^-$, shown in red).
This reveals that the GZ scar states with a particular helicity exhibit a strong overlap with the helical scar subspace of the same helicity, especially in the limit $\gamma\rightarrow 0$.
The numerical simulations are done for $S=1$ and $N=7$.
The results are independent of commensurate $q$-values ($q=4pK(\kappa)/N$).
(e) The span of the GZ scar subspace with a fixed helicity is shown as a function of $\kappa$.
The dimension of the GZ scar subspace within a given helicity sector is equal to $4NS$ for all $\kappa$, except at $\kappa = 0$, corresponding to the XXZ limit.
In the XXZ limit, the dimension of the helical scar subspace with a fixed helicity reduces to $2NS+1$. 
This mismatch in the scar-subspace dimension indicates that the standard SGA framework applicable in the XXZ limit does not admit a straightforward extension to the more general XYZ limit. 
The numerical simulations are performed for spin $S=1$.
(f) The goodness of the optimization of the SGA generator is quantified using $\log_{10}\mathcal{F}^{(2)}$, which is plotted as a function of $q$ for different spin and $\kappa$ values.
It is observed that, apart from the special points $q = 2K(\kappa)$ (marked by green oval) and $q = 4K(\kappa)$ (marked by red circles), the SGA generator $\hat{\tau}^{(2)}$ remains well optimized across all other $q$ values, with the average value $\log_{10}\mathcal{F}^{(2)}=-8.44$ represented by the black dashed line.
The different $q$ values are generated using the relation $q = 4pK(\kappa)/N$, where the system size $N$ and the parameter $p$ are varied with $p = 1,2,\ldots,N$. The system sizes considered are $N = 3,4,5,6$ for spin-$1/2$ and $N = 3,4$ for spin-$1$ systems.
}
\label{fig::ScarAlgebra}
\end{figure}

\end{widetext}

\section{Modern Algebraic Perspective}
A compelling question arises: What is the underlying algebra that leads to the emergence of such spatially periodic product states as an eigenstate of the XYZ Hamiltonian?
Significant progress has been made in understanding scar states very recently, unveiling diverse approaches to their underlying origins. 
These range from the Shiraishi-Mori construction to group-theoretical formulation of the Hamiltonian\,\cite{ShiraishiMori,GroupTheory1,GroupTheory2}, as well as insights from spectral analysis and the associated spectrum-generating algebras\,\cite{Review2, SGA2, AKLT3, Hubbard4}.
This section examines the origin of GZ scars through various lenses, focusing on the spectral properties and the algebraic structure of the Hamiltonian.

The emergence of such periodic, unentangled scar states is reflected in their spectral degeneracy, as it is formed from a superposition of many finite entangled states with the same energy.
To demonstrate this, we perform exact diagonalization on the Hamiltonian Eq.\,\ref{Eq::XYZ_Hamiltonian} for a 1D periodic chain of length $N$ and calculate the degeneracy $\mathcal{D}$ at the energy level corresponding to the GZ scars (see Appendix.\,\ref{Appendix::Sec::ScarOrigin} for more details),
\begin{equation}
    \begin{split}
    E_{GZ}= N &S^2 \cn(q,\kappa) \,\dn(q, \kappa)
    \\
    &+k^2 S^2 \sn^2(q,\kappa)\sum_n \sn(nq,\kappa) \,\sn(nq+q, \kappa).
    \end{split}
\end{equation}
The degeneracy $\mathcal{D}$ as a function of $q$-value for various system sizes and spin values is shown in Fig.\,\ref{fig::ScarAlgebra}(c).
Except for the spin-$1/2$ cases and the special $q$-values that are integer multiples of $K(\kappa)$, the degeneracy $\mathcal{D}$ follows a well-defined mathematical form: $\mathcal{D}=4NS$.
The deviations in degeneracy for the spin-$1/2$ XYZ model arise due to its integrability.
Additionally, the exceptions also occur at $q$-values that are integer multiples of $K(\kappa)$.
At these $q$-values, the model reduces to either the XY model or the isotropic Heisenberg model, with ferromagnetic or antiferromagnetic Ising interactions.
These special cases are not examined in detail in the present study, and we treat the degenerate space as having a degeneracy of $4NS$.
Next, we address the underlying algebra of the scar subspace and the relations among different scar subspaces.

\subsection{XXZ Limit: Helical-scar Subspace}
\subsubsection{Spectrum Generating Algebra (SGA) \label{SubSec::XXZ_SGA}}
In the XXZ limit, the GZ scar states reduce to the helical scar states. 
The helical scar states continue to be described by Eqs.\,\ref{Eq::GZScar} and \ref{Eq::RotationOperator}, with the Jacobi elliptic functions reducing to trigonometric sine and cosine functions in the $\kappa = 0$ limit.
The dynamics of helical states have long been of central interest in spin models, owing to their anomalous dynamical behavior\,\cite{HelixDynamics1,HelixDynamics2, HelixDynamics3, HelixDynamics4, GeneralizedSpinHelix, FelixGraphicalConstruction} and the relative ease with which they can be realized experimentally\,\cite{XXZ_chain1, XXZ_chain2, XXZ_chain3, XXZ_chain4}, due to their product state nature.
The formation of the Helical scar states can be described based on spectrum generating algebra (SGA).
The SGA formulation in the spin-$\frac{1}{2}$ XXZ limit is presented in Ref.~\cite{SGA2}. 
This approach can be generalized to arbitrary higher spin values $S$.
If $\hat{\tau}$ is the generator of SGA that generates a tower of scar states from a simple product state like $\ket{\Uparrow}$, then it must obey (see Appendix.\,\ref{Appendix::Sec::ScarOrigin}),
\begin{equation}
    \left[\hat{H}, \hat{\tau}_{\pm}\right] \ket{\mathcal{S}^{(m)}_{\pm\text{H}}}
    =
    \omega \hat{\tau}_{\pm} \ket{\mathcal{S}^{(m)}_{\pm\text{H}}},
    \label{Eq::SGA}
\end{equation}
where $\omega$ is the energy spacing between the states $\ket{\mathcal{S}^{(m)}_{\pm\text{H}}}$ and $\ket{\mathcal{S}^{(m+1)}_{\pm\text{H}}}$ in scar sub-space.
Particularly for the case of Helical scars, the states are generated as follows,
\begin{equation}
    \ket{\mathcal{S}^{(m)}_{\pm\text{H}}}=
    \frac{\hat{\tau}^m_\pm}
    {m!\sqrt{^MC_m}}
    \ket{\mathcal{S}^{(0)}_{\text{H}}}
    \;\; \text{with  }
    \hat{\tau}_\pm=\sum_n e^{\pm inq} \opS{-}{n},
    \label{Eq::XXZ_Scar_Subspace}
\end{equation}
where $\ket{\mathcal{S}^{(0)}_{\text{H}}}=\ket{\Uparrow}$ and $M=2NS$.
The scar states can be further identified using translational and total spin quantum numbers $Q$ and $s_z$, respectively,
\begin{equation}
    \ket{\mathcal{S}^{(m)}_{\pm\text{H}}}
    =
    \ket{Q=\pm mq,\, s_z=NS-m},
\end{equation}
thus forming a tower of scars shown in Fig.\,\ref{fig::ScarAlgebra}(a).
Interestingly, since the tower of scar states collapses into a degenerate energy manifold ($\omega = 0$), arbitrary superpositions of these states remain scar states.
Particularly, a special way of superposition leads to the helical scar states in terms of the tower states $\ket{\mathcal{S}^{(m)}_{\pm\text{H}}}$, as follows,
\begin{equation}
    \ket{\psi_{\pm \text{H}}}
    =\sum_{m=0}^{M} 
     \sqrt{{}^M\text{C}_m}
    \cos^{M-m} \left(\frac{\theta}{2}\right) 
    \sin^m\left(\frac{\theta}{2}\right)
    \ket{\mathcal{S}^{(m)}_{\pm \text{H}}}.
    \label{Eq::Span_of_XXZ_Scar}
\end{equation}
These helical states are identical to the GZ scars represented in Eq.\,\ref{Eq::GZScar} in the XXZ limit ($\kappa\rightarrow0$), with $\phi_n=nq$, $\theta_n=\theta$, and $\phi=0$, up to a global phase factor $\exp\left({-iSN(N+1)q/2}\right)$.
Notably, according to Eq.,\ref{Eq::Span_of_XXZ_Scar}, there are $2NS+1$ orthonormal scar states $\ket{\mathcal{S}^{(m)}_{\pm\text{H}}}$ within each helicity sector. 
Therefore, the helical scar subspace associated with a fixed helicity has dimension $2NS+1$.
The joint span of the two helicity sectors yields a total dimension of $4NS$, which matches the degeneracy of the GZ scar states, discussed at the beginning of this section and shown in Fig.\,\ref{fig::ScarAlgebra}(c).
It is noteworthy that although each helicity sector spans a space of dimension $2NS+1$, the total dimension of their joint span is $4NS$, since the two states $\ket{\Uparrow}$ and $\ket{\Downarrow}$ are shared between both the sectors.
Fig.\,\ref{fig::ScarAlgebra}(b) schematically shows the structure and interrelations between helical scar subspaces $\mathcal{H}_{\pm\text{H}}$.
It is interesting to observe that the helical scar subspaces with opposite helicities share a set of states with finite overlap, denoted by {$\mathcal{S}^{\left(p'N/f\right)}_{\pm\text{H}}$}, where $f=1$ (or $f=2$) for odd (or even) $N$.
The states corresponding to $p'=0$ and $p'=2Sf$ are the fully polarized states $\ket{\Uparrow}$ and $\ket{\Downarrow}$, respectively.
Whereas, the states with $0<p'< 2Sf$ are non-orthogonal and are unequally shared between the two helical sectors.

The algebra underlying the scar subspace for a generic spin-$S$ XXZ system is intimately connected to the algebraic Bethe ansatz structure of the integrable spin-$1/2$ limit.
It can be shown that the $B(\lambda_{p''})$(or $C(\lambda_{p''})$)-operators of the monodromy matrix for Bethe phantom roots $\lambda_{p''}=i\infty+p''\pi/n$ act as the generator $\hat{\tau}_\pm$ on the reference state  $\ket{\Uparrow}$(or $\ket{\Downarrow}$), where $p''=1,2,\ldots,n$ and $n=1,2,\ldots,N$.
Explicitly, it means,
\begin{equation}
    B(\lambda_{p''})\ket{\Uparrow}
    =
    f(\lambda_{p''})\sum_{n=1}^N 
        e^{inq}\sigma_n^- \,\,\ket{\Uparrow},
        \label{Eq::BetheGenerator}
\end{equation}
where, $\sigma_n^-=\sigma_n^x-i\sigma_n^y$ with $(\sigma_n^x,\sigma_n^y)$ as Paulli matrices.
Thus the generator $\hat{\tau}_\pm$ in Eq.\,\ref{Eq::XXZ_Scar_Subspace} is basically the generalization of the $B(\lambda_{p''})$-operator in Eq.\,\ref{Eq::BetheGenerator} in the non-intragable limit $\frac{1}{2}\sigma_n^\alpha\rightarrow\opS{\alpha}{n}$ (where $\alpha=x,y,z$).
See Appendix.\,\ref{Appendix::Sec::Integrable Limit} and reference\,\cite{PhontomBethe_PeriodicBoundary} for more details.

\subsubsection{Group Theoretical Construction}
Interestingly, when expressed in terms of Schwinger bosons, the states described in Eq.\,\ref{Eq::XXZ_Scar_Subspace} correspond to the bosonic analogs of the $\zeta$-states, described in reference\,\cite{GroupTheory2}. 
Thus, the existence of such a scar subspace suggests an underlying group-theoretical structure of the XXZ Hamiltonian, which can be expressed in the form $\hat{H}_0 + \sum_a \hat{O}_a \hat{T}_a$\,\cite{GroupTheory1, GroupTheory2}.
This decomposition consists of a simple integrable part, $\hat{H}_0$, and a set of generators, $\hat{T}_a$, of a group $G$, which satisfy the commutation relation $[\hat{H}_0, \hat{T}_a] = 0$—or, more generally, may obey a relationship involving the quadratic Casimir of the group, as discussed in Ref.\,\cite{GroupTheory1}.
Additionally, $\hat{O}_a$ denotes a set of auxiliary operators, which may or may not be constrained by the requirement of preserving the hermiticity of the Hamiltonian.
To show such a construction of the Hamiltonian, we work in the rotated basis in which the scar subspace defined in Eq.\,\ref{Eq::XXZ_Scar_Subspace} takes the form of $\zeta$-states as in Ref.\,\cite{GroupTheory2}.
This is done using the unitary operator $\prod_n \exp(inq\opS{z}{n})$ (for positive helicity sector), and this rotation simultaneously transforms the Hamiltonian (with $J_x=J_y$ and $J_z=\cos(q)$),
\begin{equation}
\begin{split}
    \hat{H}=J_x \cos(q)
    &\sum_n
    \hat{\mathbf{S}}_n
    \cdot
    \hat{\mathbf{S}}_{n+1}
    \\
    &-
    J_x \sin(q)
    \sum_n 
    \hat{z}
    \cdot
    \left(
        \hat{\mathbf{S}}_n
        \times
        \hat{\mathbf{S}}_{n+1}
    \right).
    \label{Eq::RotatedHamiltonian}
\end{split}
\end{equation}
Using Schwinger-Boson transformation $\opS{ +}{n}=\hat{c}_{n,\uparrow}^\dagger \hat{c}_{n,\downarrow}$, $\opS{-}{n}=\hat{c}_{n,\downarrow}^\dagger \hat{c}_{n,\uparrow}$, $\opS{z}{n}=\frac{1}{2}(\hat{c}_{n,\uparrow}^\dagger \hat{c}_{n,\uparrow}-\hat{c}_{n,\downarrow}^\dagger \hat{c}_{n,\downarrow})$, the Hamiltonian transformed into the following group theoretical form,
\begin{equation}
\begin{split}
    \hat{H}
    =
    &-\frac{NJ_{x} S\cos(q)}{2} 
    \\
    &
    +\frac{J_{x}\cos(q)}{4}
    \sum_n 
    \left(
    \hat{\zeta}_{n, n+1}
    \hat{\zeta}_{n+1, n}
    +
    \hat{\eta}_{n, n+1}^\dagger 
    \hat{\eta}_{n+1, n}
    \right)
    \\
    &-
    \frac{iJ_x\sin(q)}{4}
    \sum_n
    \left(
    \hat{O}^{(1)}_{n,n+1} \hat{\zeta}_{n+1,n}
    -
    \hat{O}^{(1)}_{n+1,n} \hat{\zeta}_{n,n+1}
    \right.
    \\
    &\qquad
    \left.
    +
    \hat{O}^{(2)}_{n,n+1} \hat{\varepsilon}_{n+1, n}
    -
    \hat{O}^{(2)}_{n+1, n} \hat{\varepsilon}_{n,n+1}
    \right),
    \label{Eq::H+OT_form}
\end{split}
\end{equation}
where, the first term represents $\hat{H}_0$ and the rest of the terms represent $\sum_a \hat{O}_a \hat{T}_a$.
The operators are explicitly given as,
\begin{equation}
\begin{split}
&\hat{\zeta}_{mn}
=
\hat{c}^\dagger_{m\uparrow} \hat{c}_{n\uparrow}
+
\hat{c}^\dagger_{m\downarrow} \hat{c}_{n\downarrow}
,\,
\hat{O}^{(1)}_{mn}
=
\hat{c}^\dagger_{m\uparrow} \hat{c}_{n\uparrow}
-
\hat{c}_{m\downarrow}
\hat{c}_{n\downarrow}
\\
&\hat{\eta}_{mn}
=
\hat{c}_{m\uparrow} \hat{c}_{n\downarrow}
+
\hat{c}_{m\downarrow} \hat{c}_{n\uparrow}
,\,
\hat{O}^{(2)}_{mn}
=
\hat{c}^\dagger_{m\uparrow} \hat{c}_{n\downarrow}
+
\hat{c}_{m\downarrow}^\dagger\hat{c}_{n\uparrow}
\\
&\hat{\varepsilon}_{mn}
=
\hat{c}_{m\uparrow}^\dagger \hat{c}_{n\downarrow} 
-
\hat{c}^\dagger_{m\downarrow} \hat{c}_{n\uparrow},
\end{split}    
\end{equation}
where $m\neq n$.
The operators $\hat{\zeta}$, $\hat{\eta}$, $\hat{\varepsilon}$ can be written as linear combination of $SU(N)$ generators and annihilates the bosonic $\zeta$-states,
\begin{equation}
\begin{split}
    \ket{m^\zeta}=\hat{\tau}^{\prime m} \ket{\Downarrow} 
    &\;\; \text{with  }
    \hat{\tau}^\prime=\sum_n \hat{c}_{n\uparrow}^\dagger \hat{c}_{n\downarrow},
    \\
    \text{and } \ket{\Downarrow}=\prod_{n=1}^{N} \hat{c}_{n\downarrow} \ket{0}
    \label{Eq::zeta_states}
\end{split}
\end{equation}
Comparing Eq.\,\ref{Eq::zeta_states} with Eq.\,\ref{Eq::XXZ_Scar_Subspace}, it can be observed that the $\zeta$-states are orthonormal basis $\ket{\mathcal{S}_{\text{H}}^{(m)}}$ in a rotated frame.
Moreover any operator $\hat{o}$ that annihilates the $\zeta$-state must follow a set of equations,
\begin{equation}
    \begin{split}
        \hat{o}\ket{\Downarrow}=0,\,
        [\hat{o},\hat{\tau}^\prime]\ket{\Downarrow}=0,\,
        [[\hat{o},\hat{\tau}^\prime],\hat{\tau}^\prime]\ket{\Downarrow}=0,
        \ldots
    \end{split}
\end{equation}
It can be shown that the operators $\hat{\zeta}$, $\hat{\eta}$ and  $\hat{\varepsilon}$ obey these equations.
Moreover, it is noteworthy that, to obtain the group-theoretical form of Eq.\,\ref{Eq::H+OT_form}, we add a locally non-Hermitian term, 
$iSJ_x\sin(q)\sum_n (\opS{z}{n}-\opS{z}{n+1})$, to the rotated Hamiltonian Eq.\,\ref{Eq::RotatedHamiltonian}.
A similar algebraic technique has been employed in the context of generalized helical scar states for XXC models\,\cite{GeneralizedSpinHelix}.
This transformation modifies the Hamiltonian such that its local terms are able to annihilate the scar states.
Furthermore, the existence of such an elegant scar-subspace structure is due to the quasi-$U(1)$ symmetry of the scar-subspace\,\cite{QuasiSymmetry}.
The quasi-$U(1)$ symmetry can be interpreted through the introduction of a magnetic field along the $z$-direction, resulting in equally spaced scar states that form a tower of scars, while the SGA generator serves as a ladder operator for the quasi-$U(1)$ charge.
While the quasi-$U(1)$ symmetry of the scar subspace coincides with the exact $U(1)$ symmetry of the XXZ Hamiltonian, such a correspondence does not generally persist in other systems.

\subsection{XYZ Case: GZ-scar Subspace}

In contrast to the XXZ limit, characterizing the scar subspace of the XYZ Hamiltonian presents significant challenges.
These stem, in part, from either the lack of a $U(1)$ symmetry of the XYZ model or the difficulty in identifying a quasi-$U(1)$ symmetry of the scar subspace.
Thus, rather than relying on the conventional description of the scar subspace, we explore it through several alternative approaches.
 We provide a perturbative and exact SGA approach to understand the scar subspace associated with GZ-scars.
In the perturbative SGA approach, we show how perturbatively deforming the scar-subspace in the XXZ limit still describes the scar-subspace in the XYZ limit approximately.
Whereas, in the exact SGA approach, we construct the exact SGA generator described by Eq.\,\ref{Eq::SGA} using the explicit form of the GZ states.
Third and finally, we numerically optimize an ansatz for the SGA generator and find that the generator can be accurately captured by spin operators with only two-site support.

\subsubsection{Perturbative SGA Approach}
First, we investigate whether the GZ scar states with a particular helicity and $q$-value ($q=4pK(\kappa)/N$) can still be constructed using the states within the XXZ helical-scar subspace with the same helicity and the corresponding $q$-value ($q_0=2p\pi/N$) in the $\kappa\rightarrow 0$ limit.
To address this, we evaluate the projection of the GZ scar states with positive helicity onto the helical scar subspace with positive and negative helicity, as follows,
\begin{equation}
\begin{split}
    &\mathcal{P}^{+}
    =
    \sum_{m=0}^{2NS} \left| \left\langle\Psi_{+\text{GZ}}\middle|\mathcal{S}^{(m)}_{+ \text{H}}\right\rangle \right|^2,
    \\
    &\mathcal{P}^{-}
    =
    \sum_{\substack{m=1\\m\neq \frac{p'N}{f}}}^{2NS-1} \left| \left\langle\Psi_{+\text{GZ}}\middle|\mathcal{S}^{(m)}_{- \text{H}}\right\rangle \right|^2,
\end{split}
\end{equation}
where the $q$-values used for $\ket{\Psi_{+\text{GZ}}}$ and $\ket{\mathcal{S}^{(m)}_{\pm \text{H}}}$ are $q(\kappa)=4pK(\kappa)/N$ and $q_0=q(0)=2p\pi/N$ respectively.
Moreover, the summation for $\mathcal{P}^-$ excludes some states $\ket{\mathcal{S}^{\left(\frac{pN}{f}\right)}_{\pm \text{H}}}$ with negative helicity, because of finite overlap of these states with the helical-scar subspace with positive helicity, as schematically demonstaretd in Fig.\,\ref{fig::ScarAlgebra}(b).
The projections $\mathcal{P}^{\pm}$ are plotted as a function of $\gamma$ for different values of $\kappa$ in Fig.\,\ref{fig::ScarAlgebra}(d).
It is interesting to observe that the GZ scar states with a given helicity exhibit a substantial overlap with the helical scar subspace of the same helicity, with the overlap becoming strongest in the $\gamma\rightarrow 0$ limit.

Thus, there must be a way to perturbatively derive the SGA for the GZ scars from the SGA in Eq.\,\ref{Eq::XXZ_Scar_Subspace}, defined for the helical scar subspace for the XXZ model.
We begin by performing a perturbative expansion of the XYZ Hamiltonian as a function of $\kappa$, as follows,
\begin{equation}
\begin{split}
    \hat{H} &= \sum_n \left[
    \dn(q,\kappa) \opS{x}{n} \opS{x}{n+1}
    +\opS{y}{n} \opS{y}{n+1}
    +\cn(q,\kappa)\opS{z}{n} \opS{z}{n+1}
    \right] \\
    &= \hat{H}_0+\kappa^2 \hat{H}_1
\end{split}
\end{equation}
where, 
\begin{equation}
    \begin{split}
        \hat{H}_0 &= \sum_n \left[
    \opS{x}{n} \opS{x}{n+1}
    +\opS{y}{n} \opS{y}{n+1}
    +\cos(q_0)\opS{z}{n} \opS{z}{n+1}
    \right],
    \\
    \hat{H}_1 &= -\frac{\sin^2(q_0)}{2}
    \sum_n \left[
    \opS{x}{n}\opS{x}{n+1}
    +
    \frac{\cos(q_0)}{2}
    \opS{z}{n}\opS{z}{n+1}
    \right],
    \end{split}
\end{equation}
where $q(\kappa)=4pK(\kappa)/N$ and $q_0=q(0)=2p\pi/N$. 
The first-order perturbation theory in $\kappa^2$ gives the deformed scar-subspace as,
\begin{equation}
    \ket{\mathcal{S}^{'(m)}_{\pm\text{GZ}}}
    =
    \left(
        1-\kappa^2 \hat{H}_0^{-1} \hat{H}_1
    \right)
    \ket{\mathcal{S}^{(m)}_{\pm\text{H}}}.
    \label{Eq::PerturbationTheory}
\end{equation}
However, this method of deriving the perturbed scar states is numerically challenging, as it requires computing the inverse of the many-body Hamiltonian $\hat{H}_0$. While this inversion is straightforward for a non-interacting Hamiltonian, it becomes computationally intractable in the interacting many-body case for large system sizes. 
As a result, conventional perturbation theory is not well-suited for identifying the deformed many-body scar subspace.
Surprisingly, we find that the following deformed generator is capable of reproducing the spectrum with the same accuracy as first-order perturbation theory Eq.\,\ref{Eq::PerturbationTheory},
\begin{equation}
\begin{split}
&\ket{\mathcal{S}^{''(m)}_{\pm\text{GZ}}}=\left(\hat{\tau}''\right)^m \ket{\mathcal{S}^{(0)}_{\pm\text{H}}},\;
    \hat{\tau}''=\sum_n e^{iq_n} \opS{-}{n},
    \\
    &\text{where,  } q_n=\tan^{-1}(\text{sc}(nq,\kappa)) 
\end{split}
\end{equation}
The GZ-scars can be approximately reconstructed using these perturbed states instead of the unperturbed states $\ket{\mathcal{S}^{(m)}_{\pm \text{H}}}$ in Eq.\,\ref{Eq::Span_of_XXZ_Scar} and moreover $\theta=\cos^{-1}\gamma$.
The approximation performs well as $\gamma\rightarrow 0$, whereas it becomes less accurate in the limit $\gamma\rightarrow 1$.
This behavior arises because, for $\gamma\rightarrow 0$, the GZ scar states with a given helicity have a large overlap with the corresponding helical scar subspace as shown in Fig.\,\ref{fig::ScarAlgebra}(d).
However, as will be shown next, the span of the GZ scar subspace with a given helicity is significantly larger than that of the helical scar subspace with the same helicity.
This mismatch in dimensionality leads to reduced accuracy of the approximation in the $\gamma\rightarrow 1$ limit.

\subsubsection{Exact SGA Approach \label{SubSec::ExactSGA}}
In this section, we demonstrate the way to generate an exact SGA generator using the rotation operator $\hat{\mathcal{R}}_\pm(\gamma, \phi)$.
First, we examine the span of the GZ scar subspace with a fixed helicity by generating a family of GZ scar states using the rotation operators $\left\lbrace\hat{\mathcal{R}}_{\pm}(\gamma,\phi)\right\rbrace$, obtained by varying the parameters $\gamma$ and $\phi$, and then computing the rank of the matrix formed by these states.
As shown in Fig.\,\ref{fig::ScarAlgebra}(e), the span of the GZ scar subspace with a fixed helicity equals the full scar-subspace dimension $4NS$ for all values of $\kappa$, except at $\kappa = 0$, corresponding to the XXZ limit. 
In the XXZ case, the span of the scar subspace within a given helicity sector is restricted to $2NS+1$.
This variation suggests that the SGA framework describing the XXZ helical scar subspace with a specific helicity does not admit a direct extension to the XYZ limit.
However, one can still generate an exact SGA using the rotational operators.
A family of linearly independent, non-orthonormal states can be generated by varying the parameter $\phi$ while setting $\gamma$ fixed at zero, as follows,
\begin{equation}
    \begin{split}
    \ket{\phi_{n+1}}&=\hat{\mathcal{R}}_{\pm}^{\left(\frac{2n\pi}{T}\right)}\ket{\Uparrow},\,
    \text{ for } 2n<T
    \\
    &=\hat{\mathcal{R}}_{\mp}^{\left(\frac{2n\pi}{T}\right)}\ket{\Uparrow},\,
    \text{ for } 2n\geq T
    \end{split}
    \label{Eq::LinearlyIndependentGZStates}
\end{equation}
where $T=4NS-1$, $n\in(0,1,\ldots,4NS-1)$ and $\hat{\mathcal{R}}_\pm^{(\phi)}=\hat{\mathcal{R}}_\pm(\gamma=0, \phi)$. 
From this set of states, one can generate the orthonormal set of states using the Gram–Schmidt orthonormalization protocol,
\begin{align}
    \ket{1}&=\ket{\phi_1},\,
    \ket{2}=\ket{\phi_2}-\braket{1}{\phi_2} \ket{1},
    \nonumber\\
    \ket{3}&=\ket{\phi_3}
    -\braket{1}{\phi_3} \ket{1}
    -\braket{2}{\phi_3} \ket{2},
    \cdots(4NS\text{-times}).
    \label{Eq::OrthonormalGZ_States}
\end{align}
Using these states, one can form the generating operator for SGA as follows,
\begin{equation}
    \hat{\tau}
    =
    \ket{1}\bra{2}
    +
    \ket{2}\bra{3}
    +
    \cdots
    +
    \ket{4NS-1}\bra{4NS},
    \label{Eq::ExactSGA_Generator}
\end{equation}
which can generate all the states from the reference state $\ket{4NS}$.
Because the states $\ket{1}, \ket{2},\cdots$ are made of the superposition of $O(N)$ number of product states with zero entanglement, the entanglement of the states is bounded by $O(\log(N))$, following a log-volume law of entanglement entropy (see Appendix.\,\ref{Appendix::Sec::ScarOrigin}).
Additionally, $\hat{\tau}$ in Eq.\,\ref{Eq::ExactSGA_Generator} being the generator for log-volume law entangled states can be represented using a finite bond-dimensional matrix product operator whose bond-dimension changes polynomially $\mathcal{O}(N^3)$ with the system size (see Appendix.\,\ref{Appendix::Sec::ScarOrigin}).
In this systematic manner, we recover the standard SGA from the rotation operator; however, the resulting formulation lacks the quasiparticle characteristic of the SGA\,\cite{Review1}.
Moreover, this generator is not a unique one, and
there can be infinitely many possible generators,
\begin{equation}
    \hat{\tau}^\prime
    =
    \hat{U}\hat{\tau}\hat{U}^\dagger
    +\sum_n c_n (1-\hat{P})\hat{O}_n(1-\hat{P}) \, ,
\end{equation}
where, the unitary operator $\hat{U}$ rotates the orthonormal scar states Eq.\,\ref{Eq::OrthonormalGZ_States} within the scar subspace. 
The projector $\hat{P}$ maps an arbitrary state onto the scar subspace, and the coefficients $c_n$ are complex numbers. 
Furthermore, the set of operators ${\hat{O}_n}$ forms a complete basis for operators in the spin system, consisting of all possible combinations of spin operators with varying support across the system.

\subsubsection{Numerical Optimization Approach to SGA}

Finally, we answer the question of the locality of the SGA generator by numerically optimizing an ansatz for the SGA generator.
In the previous subsection\,\ref{SubSec::XXZ_SGA}, the SGA generator for the XXZ system is shown to be made of the local operators $\opS{+}{n}$.
In this subsection, we investigate whether the SGA generator associated with the GZ scar subspace is local, and, if so, characterize its maximal support over the spin system.
We use and optimize the following ansatz of the SGA generator,
\begin{equation}
\begin{split}
    &\hat{\tau}^{(w)}
    =
    \sum_{\alpha\neq0} \sum_{n} c^{1}_{n,\alpha}  \opS{\alpha}{n}
    +
    \sum_{\alpha\neq 0, \beta\neq 0} 
    \sum_{n} 
    c^{2}_{n,\alpha,\beta}  \opS{\alpha}{n} \opS{\beta}{n+1}
    \\
    &+\sum_{\substack{\alpha\neq 0, \beta\neq 0 \\ \gamma}}
    \sum_{n} 
    c^{3}_{n,\alpha,\beta, \gamma}  
    \opS{\alpha}{n} \opS{\gamma}{n+1} \opS{\beta}{n+2}
    +
    \ldots (\text{upto } w\text{-sums}),
\end{split}    
\end{equation}
where the indices $\alpha$, $\beta$ and $\gamma$ take the following values $(x,y,z,0)$ and the index $0$ denote an identity operator.
In the above expression, the first, second, and third terms indicate the operators with single-site, two-site, and three-site support, respectively.
The maximum support size is given by $w = N$, in which case the ansatz $\hat{\tau}^{(N)}$ encompasses all possible operators of the finite spin system.
To numerically determine the correct SGA generator using the ansatz $\hat{\tau}^{(w)}$, we optimize the following function,
\begin{equation}
    \mathcal{F}^{(w)}=
    \frac{1}{(4NS)^2}
    \sum_{i=1}^{4NS} \sum_{n=0}^{4NS-1} 
    \left|\left|\,\,\left[\hat{H},\hat{\tau}^{(w)}\right](\hat{\tau}^{(w)})^n \ket{i}\,\,\right|\right|
\end{equation}
The condition $\mathcal{F}^{(w)} = 0$ implies that the generator $\hat{\tau}^{(w)}$, with support size $w$, satisfies the SGA.
Here, the states $\ket{i}$ are the orthonormal states in the GZ subspace as in Eq.\,\ref{Eq::OrthonormalGZ_States} and the symbol $||\cdot||$ defines the norm of a state.
We optimize the function for spin-$1/2$ and spin-$1$ periodic chains by varying the support $w$ using the GlobalSearch algorithm provided in MATLAB.
We observed that for the support $w=1$, the only possible solution is the trivial solution, in which case all the coefficients $c^1_{n,\alpha}$ are zero.
Whereas for the support $w=2$, the algorithm returns a non-trivial solution with a very good optimization.
The goodness of the optimization result is plotted using $\log_{10}\mathcal{F}^{(2)}$ as a function of $q$-values for spin-half and spin-one systems for different values of $\kappa$.
It is noticeable that, except for some special $q$-values such as $q=2K(\kappa)$ and $q=4K(\kappa)$, the SGA generator can be well defined using the operators with single and two site support,
\begin{equation}
    \hat{\tau}^{(2)}
    =
    \sum_{\alpha\neq0} \sum_{n} c^{1}_{n,\alpha}  \opS{\alpha}{n}
    +
    \sum_{\alpha\neq 0, \beta\neq 0} 
    \sum_{n} 
    c^{2}_{n,\alpha,\beta}  \opS{\alpha}{n} \opS{\beta}{n+1}\,.
\end{equation}

\begin{figure}[t]
\includegraphics[width=0.48\textwidth]{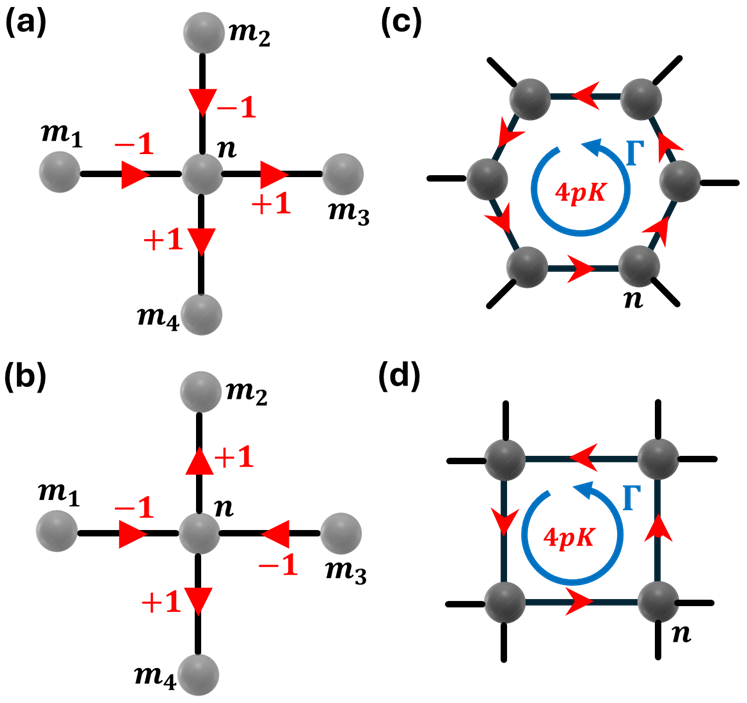}
\caption{The visual representations of the vertex rule (a), (b), and the circuit rule (c), (d).
The numbers $\pm 1$ in red and the directions with red arrowheads represent the values of $\sigma_{nm}$ associated with the edges.
Figures (a) and (b) illustrate that the number of ingoing and outgoing arrows from vertex $n$ is equal, thereby preserving the vertex rule.
Figures (c) and (d) demonstrate that along a circuit $\Gamma$, the change in variable $q_n$ in Eq.\,\ref{Eq::GZScar_Graph} is equal to $4pK(\kappa)$, where $p$ is an integer.
}
\label{fig::GraphicalRules}
\end{figure}


\section{Higher-Dimensional Realizations}
In this section, we describe how GZ scar states can be realized in higher-dimensional systems, specifically for two-dimensional lattices.
In higher dimensions, GZ scar states can, in principle, exist on any lattice.
However, their realization typically requires the Hamiltonian to be finely tuned to specific parameters, making such cases of limited practical interest.
Therefore, rather than constructing a finely tuned Hamiltonian in higher dimensions for a given GZ scar, we consider a more relevant question:
Given a generic spin-exchange Hamiltonian on a specific two-dimensional lattice that respects certain symmetries, can GZ scars emerge as eigenstates of that system?
To address this, we introduce a set of graphical rules. 
Based on these rules, we classify the types of lattices that are suitable for realizing GZ scar states.

\subsection{2D Uniform XYZ Models}

\subsubsection{Graphical Rules}

\label{Sec::TwoDimensionalScars}
\begin{figure}[t]
\includegraphics[width=0.48\textwidth]{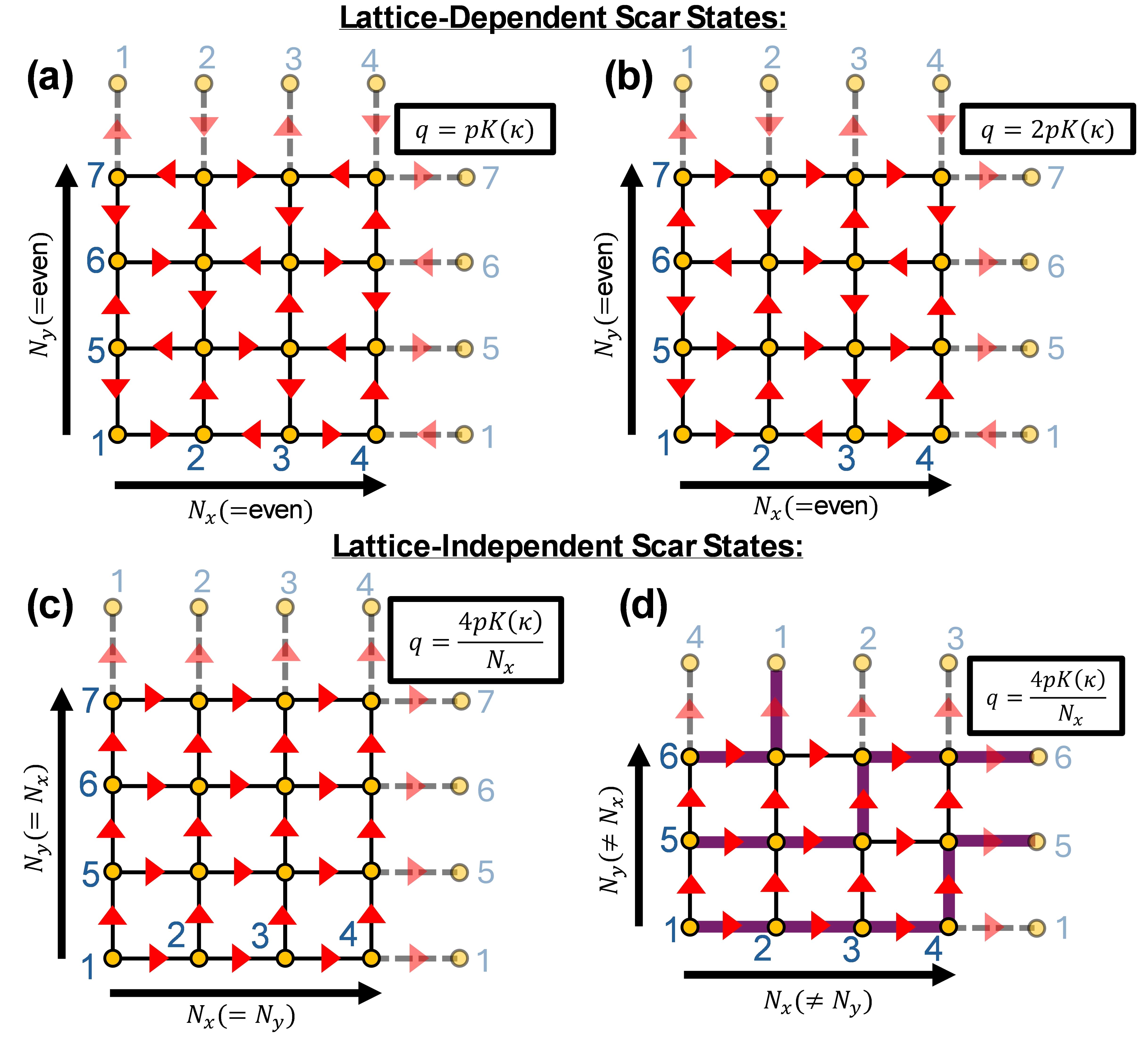}
\caption{
(a) and (b) depicts the lattice-dependent scar states which have plaquette-dependent and size-independent $q$-values given by $q=4pK(\kappa)/N_l$, where $N_l=N_p^\prime, N_p^\prime-2,\dots$. In general, $N_p$ is the number of edges of the smallest plaquette, and $p$ is an integer. 
(c) and (d) illustrates the lattice-independent scar states which have plaquette-independent and size-dependent $q$-values given by $q=4pK(\kappa)/N_\alpha$, where $\alpha$ denotes any direction $x$ or $y$ in case $N_x=N_y$, or it is the direction perpendicular to the direction of shifted boundary condition in case $N_x\neq N_y$.
}
\label{fig::SquareLattice}
\end{figure}

The GZ scar states can be naturally extended to two or more dimensions by using graphical rules. 
Inspired by Felix et al.\,\cite{FelixGraphicalConstruction}, we introduce two graphical rules for constructing scar states on arbitrarily directed simple graphs, where spins at each vertex interact through a CSSE interaction.
As described in Sec.\,\ref{Sec::Overview_of_GZ_scars}, and Fig.\,\ref{fig::1DChain}(a), a specific rotation in the reference frame transforms this kind of Hamiltonian into XYZ form.
Without loss of generality, we consider a generic spatially uniform XYZ Hamiltonian which can be parameterized as follows,

\begin{equation}
    \scalemath{0.96}{
    \hat{H}=J\sum_{n\sim m} \left[
    \dn(q,\kappa) \opS{x}{n}\opS{x}{m}
    + \opS{y}{n}\opS{y}{m}
    +\cn(q,\kappa) \opS{z}{n}\opS{z}{m}
    \right],
    }
    \label{Eq::XYZ_Hamiltonian_Graph}
\end{equation}
where the summation is over the adjacent vertices.
Scar states of this Hamiltonian are product states of spins pointing in a certain direction, such that the spins on adjacent vertices $n$ and $m$,
\begin{equation}
\begin{split}
    &\langle \opS{x}{n}\rangle=\alpha S\,\cn(q_n,\kappa),
    \,\langle \opS{x}{m}\rangle=\alpha S\,\cn(q_n-\sigma_{nm}q,\kappa),
     \\
    &\langle \opS{y}{n}\rangle=\beta S\,\sn(q_n,\kappa),
    \,\langle \opS{y}{m}\rangle=\beta S\,\sn(q_n-\sigma_{nm}q,\kappa),
     \\
    &\langle \opS{z}{n}\rangle=\gamma S\,\dn(q_n,\kappa),
    \,\langle \opS{z}{m}\rangle=\gamma S\,\dn(q_n-\sigma_{nm}q,\kappa),
\end{split}
\label{Eq::GZScar_Graph}
\end{equation}
where ($\alpha$, $\beta$, $\gamma$) are defined in Eq.\,\ref{Eq::GammaDefinition} and $\sigma_{nm}$ is binary number $\pm 1$ associated with each edge.
The states corresponding to the Eq.\,\ref{Eq::GZScar_Graph} qualifies as a GZ scars only if the following conditions are satisfied,
\begin{description}
    \item[(i) Vertex rule] For any vertex $n$, we have $\sum_m \sigma_{nm}=0$, where the summation is over all the vertices adjacent to vertex $n$.
    \item[(ii) Circuit rule] ~~Any circuit $\Gamma$ must satisfy $\sum_{(n,m)\in\Gamma} \sigma_{nm}\,q =0\,(\text{mod}\, 4K(\kappa))$, where $K(\kappa)$ is the elliptic integral of first kind.
    A circuit $\Gamma$ is a path in a graph that starts and ends at the same vertex.
\end{description}
The vertex rule can be visualized by assigning a direction to each edge, representing the value $\sigma_{nm}$ associated with it (see Fig.\,\ref{fig::GraphicalRules}(a), (b)).
The outgoing arrow from the vertex $n$ towards the adjacent vertex $m$ represents $\sigma_{nm}=+1$, while an ingoing arrow from the vertex $m$ to $n$ represents $\sigma_{nm}=-1$.
The vertex condition $\sum_{m}\sigma_{nm}=0$ literally means that the number of ingoing and outgoing arrows at each vertex must be equal.
Additionally, the underlying physics of the vertex rule holds because the $z$-component of the spin current must be zero for the scar states,
\begin{equation}
    \left\langle\dot{S}_n^z\right\rangle
    =-\alpha\beta S^2 \,\dn(q_n,\kappa) \,\sn(q,\kappa)
    \sum_{m}\sigma_{nm}=0,
\end{equation}
where the summation is over the adjacent vertices of vertex $n$.

On the other hand, the circuit rule is a consequence of the uniqueness of the spin value at each vertex. 
A closed circuit path $\Gamma$ connects vertex $n$ to itself; therefore, the total change in the spin direction along the closed path must be zero, and the required condition is,
\begin{equation}
    \sum_{(n,m)\in \Gamma}\sigma_{nm} q = 0
    \mod 4K(\kappa),
\end{equation}
where $K(\kappa)$ is the complete elliptic integral for eccentricity $\kappa$.
The visualization of the circuit rule for different graphs is depicted in Fig.\,\ref{fig::GraphicalRules}(c) and (d).

\subsubsection{Lattice Characterization \label{SubSec::LatticeCharacterization}}

\begin{table}[tb]
    \centering
    \resizebox{0.48\textwidth}{!}{
    \begin{tabular}{|c|c|c|c|}
    \toprule
          Sl. No. & Type of Lattice & \makecell{Lattice-dependent \\ GZ scars \\ (LDGZ)} & \makecell{Lattice-independent \\ GZ scars \\ (LIGZ)}  
         \tabularnewline
          \midrule
        (I) & \makecell{Odd Coordination number\\ (e.g. honeycomb, Shastry-Sutherland)} & {\color{red} \xmark} & {\color{red} \xmark}
         \tabularnewline
         \midrule
         (II) &
         \centering \makecell{Even Coordination number \\ and Plaquettes with odd edges \\ (e.g. triangular, kagome)} & {\color{teal} \cmark} & {\color{red} \xmark} 
         \tabularnewline
         \midrule
         (III) &
         \makecell{Even Coordination number\\ and Plaquettes with even edges \\ (e.g. square, Lieb)} & {\color{teal} \cmark} & {\color{teal} \cmark} or {\color{red} \xmark}
         \tabularnewline
    \bottomrule
    \end{tabular}}
    \caption{This table summarizes the possible types of scar states for different types of spatially uniform 2D lattices with centro-symmetric spin-exchange interaction. 
    The last cell contains both check and cross marks, indicating that a type-(III) lattice is a necessary condition for realizing LIGZ scars but not a sufficient one.
    }
    \label{table:lattice_types}
\end{table}

\begin{figure*}[tb]
\includegraphics[width=0.9\textwidth]{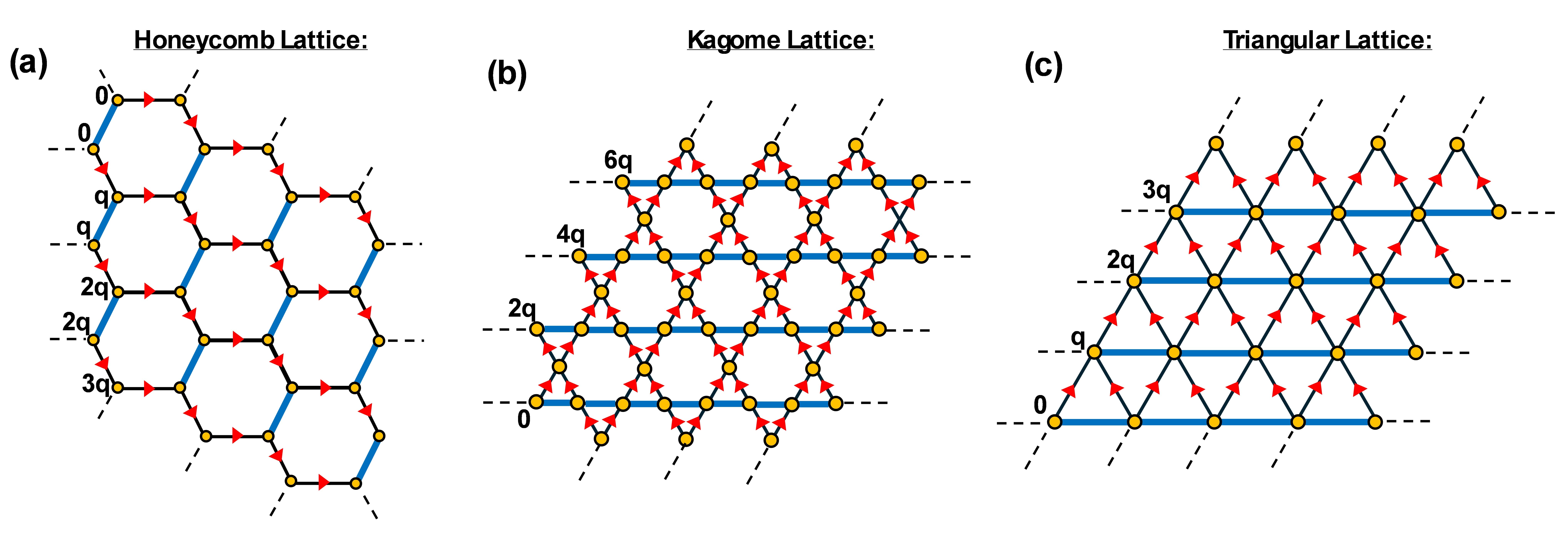}
\caption{
Demonstration of lattice-independent GZ scars on (a) honeycomb lattice, (b) kagome lattice, and (c) triangular lattice.
Spins connected by black (blue) bonds interact via a centrosymmetric (isotropic) Heisenberg exchange interaction. The red arrowhead denotes $\sigma_{nm}=+1$, whereas the $\sigma_{nm}=0$ on the blue bonds. 
The $q$-values for some of the boundary sites are shown for reference.
}
\label{fig::FrustratedModels}
\end{figure*}

In this section, we describe the construction of GZ scars on 2D lattices with NN CSSE interactions.
Firstly, not all lattices can support such scar states due to the vertex law. 
This law requires each vertex to be connected to an even number of edges. 
As a result, only certain lattices with even coordination numbers—such as square, triangular, kagome, and Lieb—are valid candidates. In contrast, lattices like honeycomb and Shastry-Sutherland do not satisfy this criterion and are therefore unsuitable.
However, the coordination number of a lattice can be modified to an even value by adding additional NN bonds, thereby making the lattice suitable for hosting GZ scar states\,(see Appendix.\,\ref{Appendix::Sec::More2DLattices}).


We examine scar states on different lattices with toroidal boundary conditions and demonstrate that scar states can be classified into two categories: lattice-dependent and lattice-independent scar states.

\begin{description}
    \item[(i) Lattice-dependent GZ (LDGZ) scars] The $q$-value of the GZ states is fixed via the underlying lattice geometry. Consequently, the Hamiltonian must be fine-tuned for each specific lattice in order to realize such GZ scars.
    
    \item[(ii) Lattice-independent GZ (LIGZ) scars] The $q$-value of the GZ states is independent of the lattice geometry. Therefore, such GZ scar states can exist in an arbitrary nearest-neighbor (NN) XYZ Hamiltonian without lattice-specific fine-tuning.

\end{description}

The $q$-values of lattice-dependent scars are limited to $q=4pK(\kappa)/N_l$, where $N_l$ can take values of $N_p, N_p-2, \dots$ and $p$ is an integer. 
Here, $N_p$ is the number of edges of a plaquette, where a plaquette is defined as a closed circuit that does not cross the periodic boundaries.
In general, the $N_p$ corresponds to the length of the smallest plaquette of a lattice.
For square, Lieb, triangular, and kagome lattices, $N_p$ takes values of $4,\,8,\,3$ and $3$ respectively (see Appendix.\,\ref{Appendix::Sec::More2DLattices}). 
Fig.\,\ref{fig::SquareLattice}(a) and (b) demonstrate two types of LDGZ scars on a square lattice with $N_l=4$ and $N_l=2$, respectively.
It is also noticeable that the $q$-values are independent of lattice dimension $N_x\times N_y$.
The LIGZ scars can exhibit a range of $q$-values that do not depend on the lattice plaquette but rather on the system's size and shape.
Figures \,\ref{fig::SquareLattice}(c) and (d) illustrate LDGZ scars on a square lattice, with (c) showing a system where the number of sites along the $x$ and $y$ directions is equal ($N_x=N_y$), and (d) depicting a system with an unequal number of sites in these directions ($N_x\neq N_y$).
The $q$-value of these scars are $q=4pK(\kappa)/ N_x$.
Noticeably, in the case of unequal lattice dimension ($N_x\neq N_y$), a twisted or shifted boundary condition is required along one of the periodic boundaries to preserve the circuit law, as demonstrated by the purple circuit in Fig.\,\ref{fig::SquareLattice}(d).
The shift in the shifted boundary condition equals $\left|N_x-N_y\right|$.
Furthermore, the $q$-value in such cases is $q=4pK(\kappa)/N_\alpha$, where $\alpha$ is the direction perpendicular to the direction of the shifted boundary condition.

A LIGZ states satisfy a strict circuit rule, $\sum_{\left(n,m\right)\in \Gamma} \sigma_{nm} q=0$ on each plaquette $\Gamma$.
Furthermore, this condition holds only when each plaquette has an even number of edges.
Thus, the triangular and kagome lattices, which contain plaquettes with three edges, are unsuitable for hosting LIGZ scars.
Although a lattice with all even plaquettes is a necessary condition for LIGZ scars, it is not sufficient, as it does not guarantee a strict circuit rule on all the plaquettes simultaneously (see the example of modified honeycomb lattice in Appendix.\,\ref{Appendix::Sec::More2DLattices}). 
Based on our findings, we have summarized which types of lattices can or cannot host GZ scar states and whether they support LDGZ or LIGZ scars in the Table.\,\ref{table:lattice_types}. 
Moreover, scar states can also be constructed in higher dimensions in a similar manner.


\subsection{Irregular 2D Lattices}

In the previous subsection\,\ref{SubSec::LatticeCharacterization}, we show that LIGZ scars are only possible in the non-frustrated lattice models with even coordination number and plaquettes with an even number of edges.
However, this statement holds only when considering a spatially uniform lattice with a single type of XYZ or CSSE interaction on NN bonds, with a Hamiltonian of the form given in Eq.\,\ref{Eq::XYZ_Hamiltonian_Graph}.
This subsection demonstrates that, given a CSSE Hamiltonian with certain bonds retaining full isotropic $SU(2)$ symmetry, it is still possible to construct LIGZ scar states on a frustrated lattice or lattices with odd coordination number.
This is demonstrated in Fig.\,\ref{fig::FrustratedModels}.
The explicit expression of the Hamiltonian of such a system is,
\begin{equation}
\begin{split}
    \hat{H} =
    &\sum_{(m,n)\in\mathcal{B}_1}\left[
    J^x
    \opS{x}{m}\opS{x}{n}
    +J^y \opS{y}{m} \opS{y}{n}
    +J^z \opS{z}{m}\opS{z}{n}
    \right]
    \\
    &+\sum_{(m,n)\in\mathcal{B}_2} 
    J^\prime\boldsymbol{\hat{S}}_m \cdot \boldsymbol{\hat{S}}_n,
\end{split}
\end{equation}
where the bonds $\mathcal{B}_1$ and $\mathcal{B}_2$ are the black and blue bonds in Fig.\,\ref{fig::FrustratedModels}, respectively.
The value of $q$ and $\kappa$ of the scars can be determined using Eq.\,\ref{Eq::q_kappa_definition}.
Moreover, the direction of the red arrow indicates $\sigma_{nm}=+1$, while its absence corresponds to $\sigma_{nm}=0$.
It can be observed that removing the blue bonds in Fig.\,\ref{fig::FrustratedModels} transforms the honeycomb, kagome, and triangular lattices into spin chains, the Lieb lattice, and the square lattice, respectively, which are the allowed lattices to have the LIGZ scar states according to the table.\,\ref{table:lattice_types}.
Therefore, the scar-state structure in the models considered here is fundamentally tied to spatially uniform NN CSSE systems supporting LIGZ scars, which are discussed in the previous subsection\,\ref{SubSec::LatticeCharacterization}.
This approach can be further extended to lattices with isotropic Heisenberg exchange interactions on trimers, as illustrated in the Appendix.\,\ref{Appendix::Sec::More2DLattices}.

\section{Discussion}

In summary, we show various aspects of GZ scars, which have been overlooked by the physics community for more than three decades and have been reintroduced in a few works very recently\,\cite{AssyemtricDecay, ExactSpinHelix}.
Although the concept of quantum scars—originally introduced by Heller in the context of single-particle systems\,\cite{Heller}—emerged around that time, their existence in many-body systems is a much more recent discovery\,\cite{PXP1, PXP2}.
The GZ scars studied in this article is especially intriguing, as these exhibit a zero entangled product and periodic eigenstate for a spatially uniform XYZ or CSSE Hamiltonian.
In this work, we aim to understand the origin of such surprising scar states through the lens of modern developments in scar construction, which have emerged over the past decade.
The existence of such scar states is rooted in the presence of a thermodynamically large degeneracy—i.e., a degeneracy that scales with the system size.
We further show that this degenerate scar subspace possesses an underlying structure, which is relatively simple to describe in the case of helical scars in the XXZ limit but significantly more complex for GZ scars in the XYZ limit.
This contrast arises from the fact that the helical scar subspace exhibits a quasi-$U(1)$ symmetry, characterized by the presence of an SGA generator that acts as a ladder operator for the quasi-$U(1)$ charges.
Due to this quasi-$U(1)$ symmetry, the helical scar subspace can be effectively described using analytically available exact SGA with a generating operator local in nature.
Moreover, the XXZ Hamiltonian admits an $\hat{H}_0+\sum_a \hat{O}_a \hat{T}_a$ group-theoretical structure which extends the depth of understanding about the presence of the helical scar subspace.
In contrast, the lack of $U(1)$ symmetry in the XYZ model renders the structure of the GZ scar subspace more challenging to characterize, owing to the absence of an SGA generator analogous to that of the XXZ model.
To address this issue, we adopt three complementary approaches—perturbative SGA, exact SGA, and numerically optimized SGA—to characterize the GZ scar subspace.
Interestingly, the span of the GZ scar subspace with a given helicity is significantly larger than that of the helical scar subspace with the same helicity.
This observation implies that the GZ scar subspace is not merely a deformed version of the helical scar subspace. 
This emphasizes that the SGA for the XXZ model cannot be directly extrapolated to obtain an SGA in the XYZ case.
Nevertheless, due to the substantial overlap between the two subspaces, the GZ states in the XYZ limit can still be well approximated within the helical scar subspace of the XXZ model, particularly in the $\gamma\rightarrow 0$ limit.
This explains the perturbative extension of the SGA from the XXZ model to the XYZ model developed in this study.
We also construct the exact SGA for the GZ scar subspace by starting from the rotation operator and employing Gram–Schmidt decomposition.
Although this procedure yields an SGA generator, it does not provide insight into the local operator structure of the generator.
Finally, to determine the local operator structure of the SGA generator, we perform a numerical optimization of an operator ansatz with variable support on the spin system. 
Our results indicate that a two-site supported SGA generator exists for all $q$ values which are integer multiples of $K(\kappa)$.

Finally, we would like to review other alternative approaches for understanding scar subspaces in the context of the GZ scar.
In the XXZ limit, the scar states reside within an $SU(2)$ subspace of the full Hilbert space, as the generator of the scar states $\hat{\tau}$ is an $SU(2)$ generator.
It is reasonable to think that in the XYZ limit, the $SU(2)$ scar subspace deforms into a $U(1)$ subspace, and thus the generator must be deformed.
However, because the XXZ SGA generator produces scar states with a fixed helicity and the dimensionality of each helicity sector in the GZ scar subspace exceeds that of the corresponding XXZ case, a deformed XXZ SGA generator cannot be constructed for the XYZ case.
Furthermore, we verified that the deformation operators constructed in the reference\,\cite{DeformedScarSpace}, which reduce the $SU(2)$ subspace to a $U(1)$ subspace, do not work in the case of GZ scars (see Appendix.\,\ref{Appendix::Sec::ScarOrigin} for more details).
Additionally, recent work has introduced systematic constructions of scar subspaces based on Fock-space caging\,\cite{FCS1,FCS2,FCS3}, which could be investigated in the future.

\begin{figure}[t]
\includegraphics[width=0.3\textwidth]{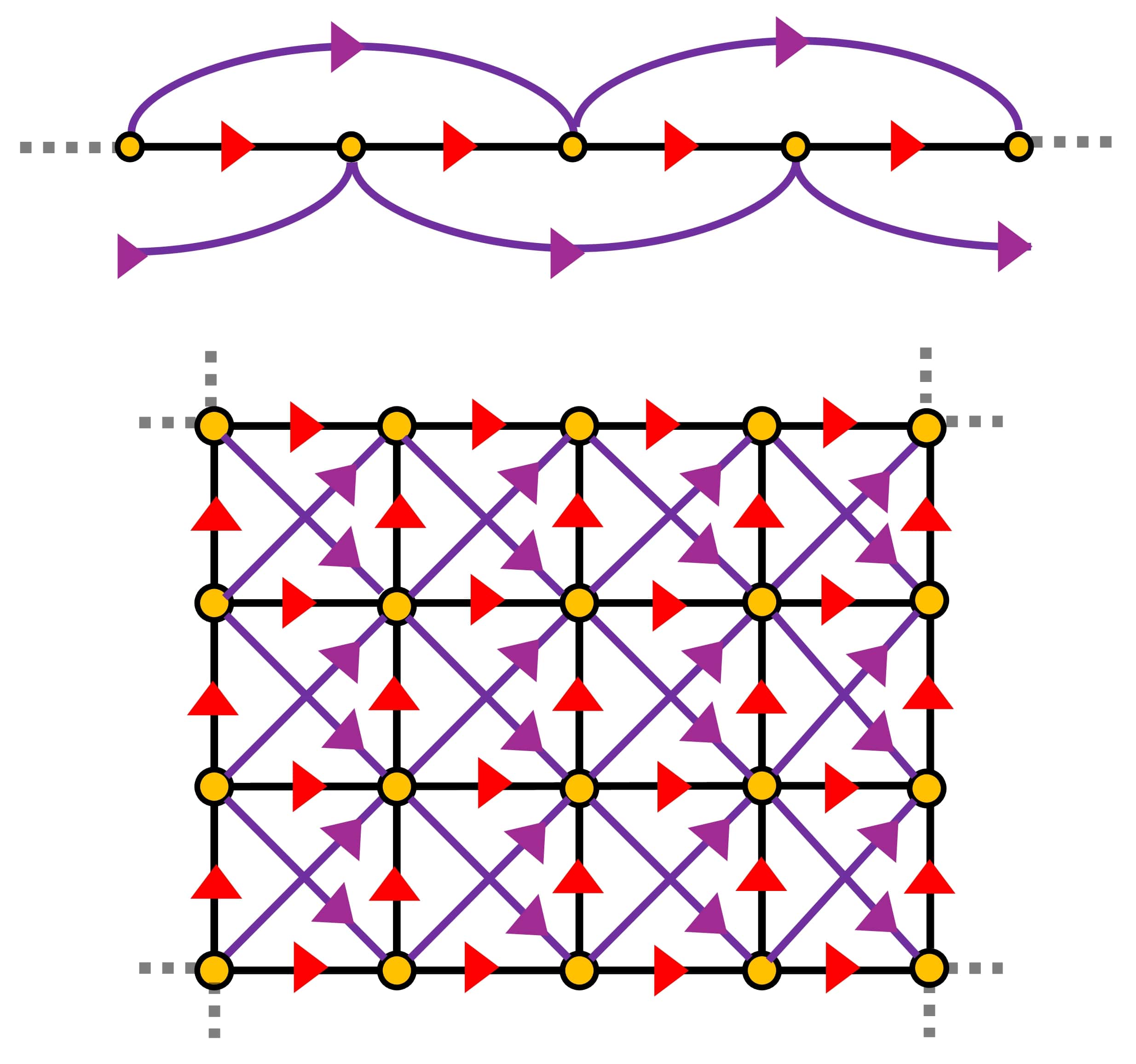}
\caption{Examples of GZ scars for fine-tuned Hamiltonians. 
The next-nearest neighbour one-dimensional spin chain (top figure) and two-dimensional square lattice (bottom figure) are shown.
Along the red (purple) arrowheads $\sigma_{nm}=+1$ ($+2$).
For such scar states, the Hamiltonian needs to be fine-tuned, which means that the interaction parameters of the Hamiltonian need to be very specific and correlated with respect to each other. In these examples, the spin-exchange interaction parameters on the next-nearest neighbour bonds have to be fixed based on the interaction parameters on the nearest-neighbour bonds.
\label{fig::MoreScarConstructions}
}
\end{figure}

We further demonstrate that GZ scars are not limited to one-dimensional lattices; they can also be constructed in higher-dimensional lattices by following specific graphical rules, referred to as the vertex rule and the circuit rule\,\cite{FelixGraphicalConstruction}.
Using these rules, we identify uniform two-dimensional lattices with nearest-neighbor CSSE interactions that are capable of hosting lattice-independent GZ scars.
Here, a lattice-independent scar is defined as a scar state whose $q$-value does not depend on the lattice type, and thus the GZ scars exist for any arbitrary NN CSSE Hamiltonian.
Lattices with an even coordination number and plaquettes having an even number of edges, such as the square and Lieb lattices, can host lattice-independent scars.
On the other hand, lattices such as the honeycomb, kagome, triangular, and Shastry–Sutherland lattices are not suitable for hosting lattice-independent scar states.
However, we also show that if certain bonds retain a fully $SU(2)$-symmetric Heisenberg exchange interaction, lattices such as the triangular, kagome, and honeycomb lattices become capable of hosting lattice-independent GZ scars.

While investigating scar-state construction on two-dimensional lattices, we try to understand the lattices that are capable of hosting lattice-independent GZ (LIGZ) scars and thus deliberately avoid fine-tuned Hamiltonians tailored to support GZ scars.
For example, a generic Hamiltonian of the following form contains GZ scars with a particular $q$-value,
\begin{equation}
\begin{split}
    \hat{H}=
    \sum_{\mathcal{B}_r}
    \sum_{(m,n)\in\mathcal{B}_r} 
    J_r
    \left[
    \dn(rq,\kappa)
    \right.&\left.\opS{x}{m}\opS{x}{n}
+\opS{y}{m}\opS{y}{n}\right.
    \\
    &\left.+\cn(rq,\kappa)\opS{z}{m}\opS{z}{n}
    \right],
\end{split}
\end{equation}
where the summation is over different bonds $\mathcal{B}_r$ and each of these bonds is associated with a $q$-value integer multiple of $q$ such that the vertex rule and the circuit rule of resultant scar states are satisfied.
We provided an example of such GZ scar constructions on such systems in Fig.\,\ref{fig::MoreScarConstructions}.
Although this construction yields a lattice-independent GZ scar state, it requires a highly fine-tuned Hamiltonian, making it of limited practical interest.
Therefore, rather than beginning with a GZ scars and determining the corresponding fine-tuned Hamiltonian, we adopt an alternative approach:
Given a spin-exchange Hamiltonian with bonds that respect either centro-symmetry or $SU(2)$ symmetry, we investigate whether lattice-independent GZ scars can emerge.
The answer is affirmative; we find that a variety of two-dimensional lattices—including square, Lieb, kagome, triangular, and honeycomb—are all capable of hosting GZ scar states under these symmetry conditions.

{\it Acknowledgments}.--- D.B. gratefully thanks Saúl Pilatowsky-Cameo, Liu Hanqing, and Daniel K. Mark for valuable discussions and insightful comments.
We also appreciate the correspondence from Alexei Zhedanov, who found our work—based on his PhD research and included in his PhD thesis—particularly interesting, and we are grateful and excited to learn about the historical development of the ideas and the relevant contemporary literature.
D.B. also acknowledges the use of computational resources provided by the Central High-Performance Computing (HPC) facility at NUS.
D.B. and W.W.H. acknowledge support from the National Research Foundation (NRF), Singapore, under the NRF Fellowship NRF-NRFF15-2023-0008. 
W.W.H. further acknowledges support from the National Quantum Office, hosted at A*STAR, under the Centre for Quantum Technologies Funding Initiative (S24Q2d0009). 



\onecolumngrid
\appendix

\section{\label{Appendix::Sec::ProofGZScar}Proof of Granovskii-Zhedanov Scar}
The proof of the existence of the periodic scar states in the XYZ Hamiltonian is provided by Granovskii and Zhedanov.
For the sake of completeness and the reader’s convenience, we reproduce the proof here.

First, we transform the Hamiltonian unitarily so that the Granovskii-Zhedanov scars become an all-up state in the rotated frame,
\begin{equation}
\ket{\Uparrow}=
    \prod_n \exp(i \opS{z}{n} \phi_n) \exp(i \opS{y}{n} \theta_n) \ket{\Psi_{GZ}}.
\end{equation}
The expressions of $\phi_n$ and $\theta_n$ are given in Eq.\,\ref{Eq::GZScar}.
This also transforms the spin operators as,

\begin{equation}
    \begin{pmatrix}
        \opS{x}{n}\\
        \opS{y}{n}\\
        \opS{z}{n}
    \end{pmatrix}
    =
    \begin{pmatrix}
        -N_n\,\sn(w_n,\kappa) & -N_n \, \sn(\overline{w}_n,\kappa) & \alpha \,\cn(nq,\kappa)
        \\
        N_n \,\cn(w_n,\kappa) & N_n \,\cn(\overline{w}_n,\kappa) & \beta \, \sn(nq, \kappa)
        \\
        iN_n & -i N_n & \gamma\,\dn(nq,\kappa)
    \end{pmatrix}
    \begin{pmatrix}
        \ops{+}{n}
        \\
        \ops{-}{n}
        \\
        \ops{z}{n}
    \end{pmatrix},
\end{equation}
where $w_n=nq+iv$ and $v$ is the incomplete elliptic integral of the first kind with elliptic modulus $k^\prime=\sqrt{1-k^2}$ and $N_n=\sqrt{1-\gamma^2\,\dn^2(nq,\kappa)}$. 
$q$ is an arbitrary parameter, and in this section, we derive its relation to the exchange coefficients.
XYZ Hamiltonian after the transformation becomes,
\begin{equation}
    \mathcal{H}^\prime
    =\sum_n 
    \left[
    \mathcal{C}_n^{--} \ops{-}{n} \ops{-}{n+1}
    +
    \mathcal{C}_n^{-+} \ops{-}{n} \ops{+}{n+1}
    +
    \mathcal{C}_n^{-z} \ops{-}{n} \ops{z}{n+1}
    +
    \mathcal{C}_n^{z-} \ops{z}{n} \ops{-}{n+1}
    +
    \text{H.C.}
    \right]
    +
    \mathcal{C}_n^{zz} \ops{z}{n} \ops{z}{n+1},
    \label{Appendix::Eq::RoatatedFrameHamiltonian}
\end{equation}
where the coefficients are given by,
\begin{gather}
    \mathcal{C}_n^{--}
    =
    N_n N_{n+1}\left[
    J_x\,\sn(\overline{w}_n,\kappa) \,\sn(\overline{w}_{n+1},\kappa)
    +
    J_y\,\cn(\overline{w}_n,\kappa) \,\cn(\overline{w}_{n+1},\kappa)
    -Jz
    \right]
    \nonumber\\
    \mathcal{C}_n^{-+}
    =
    N_nN_{n+1}\left[
    J_x\,\sn(\overline{w}_n)\,\sn(w_{n+1})
    +J_y\,\cn(\overline{w}_n) \,\cn(w_{n+1})
    +J_z
    \right]
    \nonumber\\
    \mathcal{C}_n^{-z}
    =
    -N_n\left[
    J_x\alpha\,\sn(\overline{w}_n) \,\cn(nq+q)
    -
    J_y \beta \,\cn(\overline{w}_n) \,\sn(nq+q)
    +
    iJ_z\gamma \,\dn(nq+q)
    \right]
    \nonumber \\
    \mathcal{C}_n^{z-}
    =
    -N_{n+1} \left[
    J_x\alpha \,\cn(nq)\,\sn(\overline{w}_{n+1})
    -
    J_y \beta \,\sn(nq)\,\cn(\overline{w}_{n+1})+iJ_z\gamma\,\dn(nq)
    \right]
    \nonumber \\
    \mathcal{C}_n^{zz}
    =
    \left[
    J_x \alpha^2 \,\cn(nq)\,\cn(nq+q)
    +
    J_y \beta^2 \,\sn(nq) \,\sn(nq+q)
    +
    J_z \gamma^2 \,\dn(nq)\,\dn(nq+q)
    \right].
    \label{Appendix::Eq::AllCoefficients}
\end{gather}
If the state $\ket{\Psi}_{GZ}$ is a scar state of the XYZ Hamiltonian, then the following eigenvalue equation must be satisfied,
\begin{flalign}
    &\;\;\;\;\;\;\;\;\;\;\;
    \mathcal{H}^\prime \ket{\Uparrow}=E\ket{\Uparrow}&&
    \nonumber \\
    &\implies
    \sum_n
    \left[
    \left\lbrace
    \mathcal{C}_n^{--} \ops{-}{n} \ops{-}{n+1}
    +
    \mathcal{C}_n^{-+} \ops{-}{n} \ops{+}{n+1}
    +
    \mathcal{C}_n^{-z} \ops{-}{n} \ops{z}{n+1}
    +
    \mathcal{C}_n^{z-} \ops{z}{n} \ops{-}{n+1}
    +
    \text{H.C.}
    \right\rbrace
    +
    \mathcal{C}_n^{zz} \ops{z}{n} \ops{z}{n+1}
    \right]
    \ket{\Uparrow}
    = E\ket{\Uparrow}&&
    \nonumber \\
    &\implies
    \sum_n
    \left[
    \mathcal{C}_n^{--} 
    \ops{-}{n}
    \ops{-}{n+1}
    +
    \mathcal{C}_{n}^{-z} 
    \ops{-}{n}
    \ops{z}{n+1}
    +
    \mathcal{C}_{n}^{z-}
    \ops{z}{n}
    \ops{-}{n+1}
    +
    \mathcal{C}_n^{zz}
    \ops{z}{n}
    \ops{z}{n+1}
    \right]
    \ket{\Uparrow}
    =E\ket{\Uparrow}&&
    \nonumber \\
    &\implies
    \sum_n
    \left[
    \mathcal{C}_n^{--} 
    \ops{-}{n}
    \ops{-}{n+1}
    +S
    \mathcal{C}_{n}^{-z} 
    \ops{-}{n}
    +S
    \mathcal{C}_{n}^{z-}
    \ops{-}{n+1}
    +S^2
    \mathcal{C}_n^{zz}
    \right]
    \ket{\Uparrow}
    =E\ket{\Uparrow}&&
    \nonumber \\
    &\implies
    \sum_n
    \left[
    \mathcal{C}_n^{--} 
    \ops{-}{n}
    \ops{-}{n+1}
    +S
    \left(
    \mathcal{C}_{n}^{-z} 
    +
    \mathcal{C}_{n-1}^{z-}
    \right)
    \ops{-}{n}
    +S^2
    \mathcal{C}_n^{zz}
    \right]
    \ket{\Uparrow}
    =E\ket{\Uparrow}.&&
\end{flalign}
To be a valid eigen equation, the following conditions must be satisfied,
\begin{equation}
    \mathcal{C}_n^{--}=0 
    \;\;\;\;\text{and}\;\;\;\;\;
    \mathcal{C}_{n}^{-z} 
    +
    \mathcal{C}_{n-1}^{z-}=0.
\end{equation}
This condition holds if,
\begin{equation}
    \dn(q,\kappa)=\frac{J_x}{J_y},\, \cn(q,\kappa)=\frac{J_z}{J_y},\,
\end{equation}
and it can be proved by using the following identities,
\begin{gather}
\cn(a,\kappa)\,\cn(b,\kappa)+\dn(a-b,\kappa)\sn(a,\kappa)\sn(b,\kappa)-\cn(a-b,\kappa)=0
    \nonumber \\
    \dn(a-b,\kappa)\sn(a,\kappa)\cn(b,\kappa)-
    \cn(a,\kappa)\sn(b,\kappa)
    -\dn(a,\kappa)\sn(a-b,\kappa)=0.
\end{gather}

\section{\label{Appendix::Sec::ThetaState} Connection to $\vartheta$-state}
Motivated by the existence of spin-helix eigenstates in spin-$1/2$ XYZ systems~\cite{IntegrableXYZ1,IntegrableXYZ2,IntegrableXYZ3}, Mingchen Zheng et al. recently rediscovered analogous states for spin-$S$ XYZ chains~\cite{ExactSpinHelix}. 
We refer to these states as $\vartheta$-states, as they are constructed using Jacobi theta functions. 
Although Jacobi theta functions are known to be closely related to Jacobi elliptic functions, the connection between the $\vartheta$-states and the Granovskii–Zhedanov (GZ) states is not immediately transparent.
In this appendix, we explicitly demonstrate the equivalence between these two constructions.
Firstly, we define $\theta_\alpha(\eta)=\vartheta_\alpha(\pi \eta, e^{i\pi\tau})$, $\tilde{\theta}_\alpha(\eta)=\vartheta_\alpha(\pi \eta, e^{2i\pi\tau})$, where $\vartheta_\alpha(u,q)$ represents four Jacobi elliptic functions with $\alpha=1,2,3,4$.
Using this notation, the XYZ Hamiltonian can be parameterized in terms of Jacobi theta functions as,
\begin{equation}
    J_x=\frac{\theta_3(\eta)}{\theta_3(0)},\,
    J_y=\frac{\theta_4(\eta)}{\theta_4(0)},\,
    J_z=\frac{\theta_2(\eta)}{\theta_2(0)}
\end{equation}
These XYZ coefficients are connected with the XYZ coefficients in terms of Jacobi elliptic functions in Eq.\,\ref{Eq::q_kappa_definition}, if,
\begin{equation}
    q=\pi \eta (\theta_3(0))^2,\,\, \kappa=\left(\frac{\theta_2(0)}{\theta_3(0)}\right)^2.
    \label{Eq::parameter1}
\end{equation}
Furthermore, the $\vartheta$-state corresponding to the XYZ Hamiltonian is given by,
\begin{equation}
    \ket{\Psi_\vartheta}=\prod_n 
    \exp\left(-i\beta_n\opS{z}{n}\right)
    \exp\left(-i \gamma_n \opS{y}{n}\right)
    \ket{\Uparrow},
\end{equation}
where,
\begin{equation}
\gamma_n=2\tan^{-1}\left|\frac{\tilde{\theta}_4(n\eta+u)}{\tilde{\theta}_1(n\eta+u)}\right|,\,
    \beta_n=\arg \left(\frac{\tilde{\theta}_4(n\eta+u)}{\tilde{\theta}_1(n\eta+u)}\right),
\end{equation}
where $u$ is a complex free parameter.
The free parameters $\gamma$ and $\phi$ of the GZ state must be connected with the free parameter $u$ of $\vartheta$-state.
We determine that,
\begin{equation}
    \phi=\pi \text{Re}[u] \left(\theta_3(0)\right)^2,\,
    \gamma=-\frac{\cos (\gamma_0)}{\cos(\gamma_0^\prime)},
    \label{Eq::parameter2}
\end{equation}
where, $\gamma_0^\prime=2\tan^{-1}\left|\frac{\tilde{\theta}_4(\text{Re}[u])}{\tilde{\theta}_1(\text{Re}[u])}\right|$.
Using the parameters $q$, $\kappa$, $\phi$, $\gamma$ defined in Eq.\,\ref{Eq::parameter1} and Eq.\,\ref{Eq::parameter2}, we determine that the GZ state and the $\vartheta$-state are equivalent up to an overall global phase factor, as follows,
\begin{equation}
    \ket{\Psi_{-\text{GZ}}}=e^{i\frac{3\pi}{4}} \ket{\Psi_{\vartheta}}.
\end{equation}

\section{\label{Appendix::Sec::ScarOrigin} Additional Derivations and Discussions related to the SGA and Degeneracy}

\subsection{Degeneracy of Scar Subspace}
\begin{figure}[t]
\includegraphics[width=\textwidth]{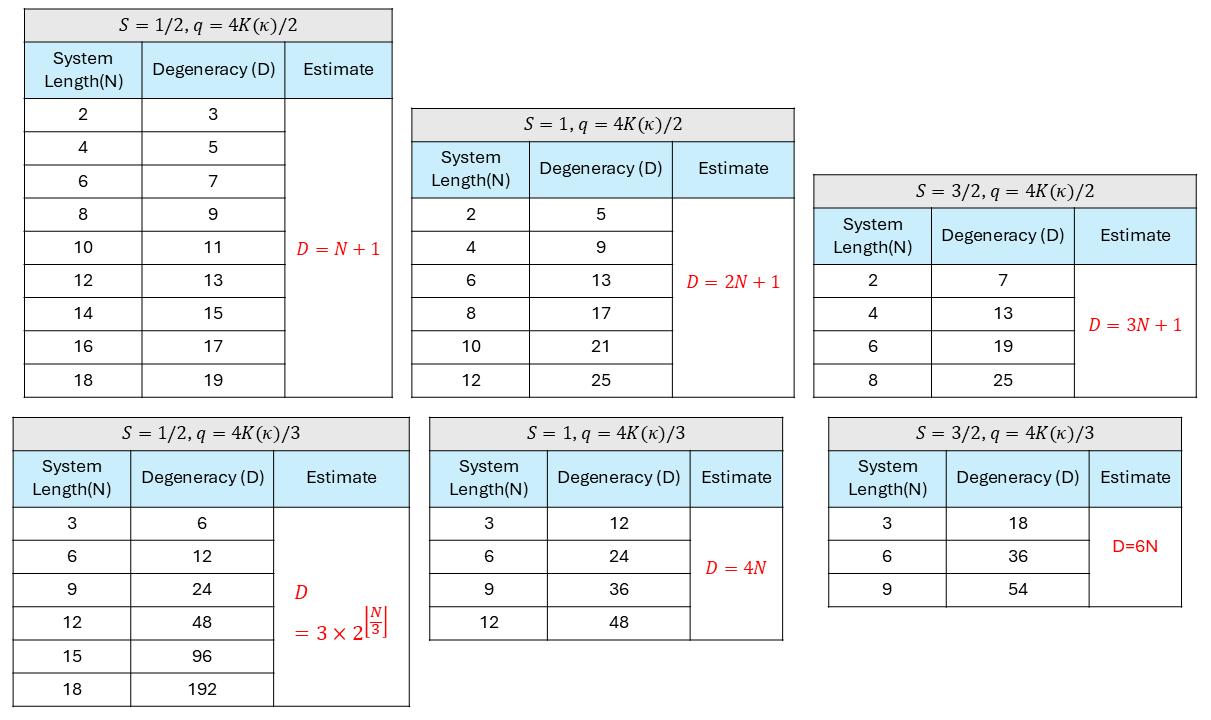}
\caption{
Degeneracy of the energy level corresponding to the GZ scar, calculated using the exact diagonalization technique for several spin and $q$-values.
 It is observed that the degeneracy generally increases linearly with the system size, except for $S=1/2$ or $q=4K(\kappa)/2$, where the integrability of the model causes deviations from this trend.
}
\label{Appendix::fig::ScarDegeneracy}
\end{figure}

To understand the origin of the GZ scar, we study the spectral degeneracy at the energy of the GZ scar,
\begin{equation}
    E_{GZ}= N S^2 \cn(q,\kappa) \,\dn(q, \kappa)
    +k^2 \sn^2(q,\kappa)\sum_n \sn(nq,\kappa) \,\sn(nq+q, \kappa).
\end{equation}
We perform exact diagonalization of the spin system using the QuSpin Python package\,\cite{QuSpin}. 
By leveraging the translational and spin-flip symmetries, we aim to diagonalize the largest possible system size.
The numerical data are tabulated in Fig.\,\ref{Appendix::fig::ScarDegeneracy}, and we further analyze the degeneracy pattern as a function of system size, as presented in the table.
Apart from special cases such as $S=1/2$ and $q$-values that are integer multiples of $K(\kappa)$, the degeneracy of the scar subspace is observed to be $4NS$.
This finding further supports the analysis presented in the main text.

\subsection{Spectrum Generating Algebra (SGA) in XXZ limit}
Moreover, in the main text, we described the scar state in the XXZ limit using the spectrum generating algebra (SGA). 
In this appendix, we elaborate on it.
Suppose $\hat{H}$ is the XXZ Hamiltonian, then it satisfies the following commutation relation with the generator $\hat{\tau}$,
\begin{equation}
    \left[\hat{H},\hat{\tau}\right]=\hat{\Lambda},
    \;\;\text{with}\;\;
    \hat{\tau}=\sum_n e^{inq} \opS{-}{n}
    \;\;\text{and}\;\;
    \hat{\Lambda}=i\sin(q)\sum_n e^{inq} \opS{-}{n} \left[\opS{z}{n+1}-\opS{z}{n-1}\right].
\end{equation}
Additionally, it can be shown that the following relations also hold,
\begin{equation}
    \left[\hat{\tau}, \hat{\Lambda}\right]=0,
    \;\;
    \hat{\Lambda}\ket{\Uparrow}=0,
    \;\;
    \hat{H}\ket{\Uparrow}=E_{GZ} \ket{\Uparrow}
\end{equation}
Using the above equations we can show that the state $\ket{\mathcal{S}^{(m)}}=\hat{\tau}^m\ket{\Uparrow}$ is also eigenstate of $\mathcal{H}$, as follows,
\begin{align*}
    \hat{H}\ket{\mathcal{S}^{(m)}}
    &=\hat{H} \hat{\tau}^m \ket{\Uparrow}
    \\
    &=\left[\hat{\tau}\hat{H}+\hat{\Lambda}\right] \hat{\tau}^{m-1}\ket{\Uparrow}
     \\
    &=\left[\hat{\tau}\hat{H}\hat{\tau}^{m-1}+\hat{\tau}^{m-1}\hat{\Lambda}\right] \ket{\Uparrow}
    \\
    &=\hat{\tau}\hat{H}\hat{\tau}^{m-1}\ket{\Uparrow} 
    \qquad\qquad\qquad
    \left[\because\hat{\Lambda}\ket{\Uparrow}=0\right]
    \\
    &\vdots
    \\
    &=\hat{\tau}^m \hat{H}\ket{\Uparrow}
    \\
    &=E_{GZ} \hat{\tau}^m \ket{\Uparrow}
\end{align*}

\subsection{Entanglement of the Scar Subspace and Bond Dimension of the SGA Generator}
In this section, we derive the log-volume law of entanglement entropy for the scar subspace and the maximum bond dimension of the SGA generator in MPO representation, which are discussed in the subsection.\,\ref{SubSec::ExactSGA}. 
The linearly independent GZ states generated using rotation operators in Eq.\,\ref{Eq::LinearlyIndependentGZStates} are product states and thus can be written in the following explicit form $\ket{\phi_n}=\ket{L_n}\ket{R_n}$, where $\ket{L_n}$ ($\ket{R_n}$) is the state of the left (right) sub-system.
The orthogonal states generated using the Gram-Schmidt decomposition can be written as,
\begin{equation}
    \ket{n}
    =\sum_m c_m \ket{\phi_m}
    =\sum_m c_m \ket{L_m} \ket{R_m}.
\end{equation}
The state $\ket{n}$ would be maximally entangled if all the states in the superposition equally participate, which means that $\ket{n}=\frac{1}{\sqrt{4NS}}\sum_{m=1}^{4NS} e^{i\chi_m} \ket{L_m}\ket{R_m}$ and states of the left (right) subsystem $\left\lbrace\ket{L_m}\right\rbrace$ ($\left\lbrace\ket{R_m}\right\rbrace$) form an orthonormal set of states.
Then, the probability of each of the $4NS$ configurations is $p_m=1/4NS$.
Therefore, the Von Neumann entanglement entropy is given by,
\begin{equation}
    S_{\text{vN}}=\sum_m p_m \log (1/p_m)
    =\log(4NS)=\mathcal{O}(\log(N)).
\end{equation}
This proves the GZ scar subspace follows a log-volume law. Next, we will show that the bond dimension of the SGA generator is $\mathcal{O}(N^3)$.

Due to the log-volume law of the scar subspace, the orthonormal scar states in Eq.\,\ref{Eq::OrthonormalGZ_States} can be represented as the matrix product state (MPS) as follows,
\begin{equation}
    \ket{n}=\sum_{\left\lbrace s_i\right\rbrace}
    \text{Tr}\left[ A_n^{s_1} A_n^{s_1}\cdots A_n^{s_N}\right]
    \ket{s_1 s_2 \cdots s_N},
\end{equation}
where $s_i$ denotes the local quantum state at $i$-th site and $A_n^{s_i}$ are the matrices of dimension (or bond-dimension) $\mathcal{O}(N)$. 
Thus, each term of the SGA generator in Eq.\,\ref{Eq::ExactSGA_Generator} can be written as a matrix product operator (MPO) as,
\begin{equation}
    \ket{n}\bra{n+1}
    =
    \sum_{\left\lbrace s_i \right\rbrace}
    \sum_{\left\lbrace t_i \right\rbrace}
    \text{Tr}\left[
    \left(A_n^{s_1} \otimes (A_{n+1}^{t_1})^*\right)
    \left(A_n^{s_2} \otimes (A_{n+1}^{t_2})^*\right)
    \cdots
    \left(A_n^{s_N} \otimes (A_{n+1}^{t_N})^*\right)
    \right]
    \ket{s_1 s_2 \cdots s_N}
    \bra{t_1 t_2 \cdots t_N}.
\end{equation}
Thus, the operators $\ket{n}\bra{n+1}$ can be represented as MPO with bond dimension $\mathcal{O}(N^2)$.
The SGA generator in the MPO representation is,
\begin{equation}
    \hat{\tau}
    =
    \sum_{n=1}^{4NS-1}
    \ket{n}\bra{n+1}
    =
    \sum_{\left\lbrace s_i\right\rbrace}
    \sum_{\left\lbrace t_i\right\rbrace}
    \text{Tr}
    \left[
    W^{s_1,t_1} W^{s_2,t_2} \cdots W^{s_N, t_N}
    \right]
    \ket{s_1 s_2 \cdots s_N}
    \bra{t_1 t_2 \cdots t_N},
\end{equation}
where,
\begin{equation*}
    W^{s_i, t_i}
    =
    \left(A_1^{s_i} \otimes (A_{2}^{t_i})^*\right)
    \oplus
    \left(A_2^{s_i} \otimes (A_{3}^{t_i})^*\right)
    \oplus
    \cdots
    \oplus
    \left(A_{4NS-1}^{s_i} \otimes (A_{4NS}^{t_i})^*\right).
\end{equation*}
Because there are $4NS-1$ matrices of dimension $\mathcal{O}(N^2)$ is in the direct sum, the dimension of the matrix $W^{s_i, t_i}$ is $\mathcal{O}(N^3)$.
Therefore, the SGA generator $\hat{\tau}$ can be represented as a matrix product operator with bond dimension $\mathcal{O}(N^3)$.

\subsection{Deformation Operators}
We check for the suitable spin-$1/2$ deforming operators that deform the $SU(2)$ scar subspace to $U(1)$ scar subspace.
The type-1 and the type-2 deforming operators are defined by\,\cite{DeformedScarSpace},
\begin{equation}
    \begin{split}
        W_{\text{type-1}}=a\ket{\uparrow}\bra{\uparrow} + b\ket{\downarrow}\bra{\downarrow},
        \,\,
        W_{\text{type-2}}=
        \begin{pmatrix}
            a\ket{\uparrow}\bra{\uparrow} + b\ket{\downarrow}\bra{\downarrow} & 
            e\hat{\sigma}^+ 
            \\
            f\hat{\sigma}^- 
            &
            c\ket{\uparrow}\bra{\uparrow} + d\ket{\downarrow}\bra{\downarrow}
        \end{pmatrix}.
    \end{split}
\end{equation}
These deforming operators leave the fully-polarized state $\ket{\Uparrow}$ unchanged.
Since $\ket{\Uparrow}$ is not a basis state of the scar subspace in the XYZ case, these operators are not suitable for deforming the scar subspace.
Thus the only remaining option for the deforming operator is the type-3 deforming operator,
\begin{equation}
    W_{\text{type-3}}=
    \begin{pmatrix}
        a\ket{\uparrow}\bra{\uparrow} & c\hat{\sigma}^+ 
        \\
        d\ket{\uparrow}\bra{\uparrow} 
        &
        b\ket{\downarrow}\bra{\downarrow}.
    \end{pmatrix}
\end{equation}
This is also not a suitable deforming operator as it annihilates the $\ket{\Uparrow}$ state.
However, the deformation must happen adiabatically such that there must be a scar state in the XYZ case corresponding to $\ket{\Uparrow}$ state in the XYZ limit.
Thus, there are no suitable deforming operators described in Ref.\,\cite{DeformedScarSpace} that deforms the $SU(2)$ scar subspace of the XXZ Hamiltonian to describe the GZ scar subspace as a $U(1)$ scar subspace.

\section{\label{Appendix::Sec::Integrable Limit} Integrable Limit}

\subsection{Spin-1/2 XXZ}
The R-matrix for the XXZ spin chain is given as,
\begin{equation}
    R(\lambda)
    =
    \begin{pmatrix}
        a(\lambda) & 0 & 0 & d(\lambda)\\
        0 & b(\lambda) & c(\lambda) & 0\\
        0 & c(\lambda) & b(\lambda) & 0\\
        d(\lambda) & 0 & 0 & a(\lambda)
    \end{pmatrix},
\end{equation}
where 
$
a(\lambda)=\rho \sin(\lambda + q ),\,
b(\lambda)=\rho \sin(\lambda ),\,
c(\lambda)=\rho \sin(q),\,
d(\lambda )=0
$
with $\rho$,$\lambda$, and $q$ as the normalization constant, rapidity, and the Hamiltonian parameter, respectively.
It can be shown that the local XXZ Hamiltonian term can be expressed as,
\begin{equation}
    h_{n,n+1}=P_{n,n+1} \left.\frac{dR_{n,n+1}(\lambda)}{d\lambda}\right|_{\lambda=0}
    =\rho \left[
    \sigma_n^x \sigma_{n+1}^x
    + \sigma_n^y \sigma_{n+1}^y
    + cos(q) 
    \left(
    \sigma_n^z \sigma_{n+1}^z
    +
    \mathbb{I}_{n,n+1}
    \right)
    \right],
\end{equation}
where, $P_{n,n+1}$ represent the permutation operator for site $n$ and $n+1$ and matrix representation of the operator is as follows,
\begin{equation}
    P=
    \begin{pmatrix}
        1 & 0 & 0 & 0\\
        0 & 0 & 1 & 0\\
        0 & 1 & 0 & 0\\
        0 & 0 & 0 & 1
    \end{pmatrix}.
\end{equation}
It is interesting to note that $R(0)\propto P$. This is known as the regularity condition, which assures that the Hamiltonian corresponding to the R-matrix is local.

The Lax-operator is given by $L_n(\lambda)=R_{0,n}(\lambda)$, where the index $0$ and $n$ represent an ancilla and physical spin-$1/2$ site.
The Lax operator can be represented as,
\begin{equation}
    L_n(\lambda)=
    \begin{pmatrix}
        \frac{a(\lambda)+b(\lambda)}{2} \mathbb{I}_n+\frac{a(\lambda)-b(\lambda)}{2} \sigma_n^z
        &
        \frac{d(\lambda)+c(\lambda)}{2} \sigma_n^x
        +i\frac{d(\lambda)-c(\lambda)}{2} \sigma_n^y
        \\
        \frac{d(\lambda)+c(\lambda)}{2} \sigma_n^x
        -i\frac{d(\lambda)-c(\lambda)}{2} \sigma_n^y
        &
        \frac{a(\lambda)+b(\lambda)}{2} \mathbb{I}_n-\frac{a(\lambda)-b(\lambda)}{2} \sigma_n^z
    \end{pmatrix}
\end{equation}
The monodromy matrix can be derived from the symbolic matrix multiplication (where the matrices $\mathbb{I}$, $\sigma_n^\alpha$ are treated as a scalar symbols instead of matrix operators),
\begin{align}
    T(\lambda)
    &=
    L_N(\lambda) L_{N-1}(\lambda) \ldots L_1(\lambda)
    \nonumber \\
    &=\begin{pmatrix}
        A(\lambda) & B(\lambda) \\
        C(\lambda) & D(\lambda)
    \end{pmatrix}.
\end{align}

We are particularly interested in deriving the $B(\lambda)$-operator of the monodromy matrix at the Bethe Phantom root $\lambda_p=i\lambda_\infty+\frac{p\pi}{n}$.
The matrix elements at $\lambda_p$ becomes,
\begin{equation}
    a(\lambda_p)=-\frac{\rho}{2i} e^{\lambda_\infty} e^{-i\left(q +\frac{p\pi}{n}\right)},\,
    b(\lambda_p)=-\frac{\rho}{2i} e^{\lambda_\infty} e^{-i\frac{p\pi}{n}},\,
    c(\lambda_p)=\rho \sin(q),\,
    d(\lambda_p)=0.
\end{equation}
Because the reference state is $\ket{\Uparrow}$, one can generate all the eigenstates of the XXZ Hamiltonian by operating the $B(\lambda)$ operator on this state.
The Lax operator at the Bethe phantom root operated on the reference state is given by,
\begin{equation}
    L_n(\lambda_p)
    =
    \begin{pmatrix}
        a(\lambda_p) \mathbb{I}_n
        &
        c(\lambda_p) \sigma_n^-
        \\
        0
        &
        b(\lambda_p) \mathbb{I}_n
    \end{pmatrix}
    \ket{\uparrow_n}
\end{equation}
Thus, the monodromy matrix at the Bethe phantom root acting on the reference state is given by,
\begin{align}
    T(\lambda_p)\ket{\Uparrow}
    &=
    L_N(\lambda_p)L_{N-1}(\lambda_p)\cdots
    L_1(\lambda_p)
    \ket{\Uparrow}
    \nonumber\\
    &=
    \begin{pmatrix}
        a(\lambda_p) \mathbb{I}_N
        &
        c(\lambda_p) \sigma_N^-
        \\
        0
        &
        b(\lambda_p) \mathbb{I}_N
    \end{pmatrix}
    \begin{pmatrix}
        a(\lambda_p) \mathbb{I}_{N-1}
        &
        c(\lambda_p) \sigma_{N-1}^-
        \\
        0
        &
        b(\lambda_p) \mathbb{I}_{N-1}
    \end{pmatrix}
    \cdots
    \begin{pmatrix}
        a(\lambda_p) \mathbb{I}_1
        &
        c(\lambda_p) \sigma_1^-
        \\
        0
        &
        b(\lambda_p) \mathbb{I}_1
    \end{pmatrix}
    \ket{\Uparrow}
\end{align}
From the above equation the $B(\lambda_p)$-operator can be extracted as,
\begin{align}
    B(\lambda_p)\ket{\Uparrow}
    &=
    c(\lambda_p) 
    \left[
        a(\lambda_p)^{N-1} \sigma_1^-
        +
        a(\lambda_p)^{N-2} b(\lambda_p) \sigma_2^-
        +
        a(\lambda_p)^{N-3} b(\lambda_p)^2 \sigma_3^-
        +\cdots(N\text{-terms})
    \right]
    \ket{\Uparrow}
    \nonumber \\
    &=
    -\left(\frac{e^{\lambda_\infty}}{2i}\right)
    e^{-i\left((N-1)\frac{p\pi}{N}\right)}
    \sin(q)
    e^{-i\left(\frac{N}{2}+1\right)q}
    \left[
    e^{iq} \sigma_1^-
    +
    e^{i2q} \sigma_2^-
    +
    e^{i3q} \sigma_2^-
    +
    \cdots(N\text{-terms})
    \right]
    \nonumber \\
    &=
    f(\lambda_p)\sum_{n=1}^N 
        e^{inq}\sigma_n^- \,\,\ket{\Uparrow}.
\end{align}
Thus, we have recovered Eq.\,\ref{Eq::BetheGenerator} in the main text.

\subsection{Spin-1/2 XYZ}
In this section, we show that the Bethe algebra structure of the spin-$1/2$ XYZ model possesses a similar structure to the spin-$1/2$ XXZ model.
The details of this section are based on the reference\,\cite{BetheXYZ}.
The R-matrix for the XYZ spin chain is,
\begin{equation}
    R(\lambda)
    =
    \begin{pmatrix}
        a(\lambda) & 0 & 0 & d(\lambda)\\
        0 & b(\lambda) & c(\lambda) & 0\\
        0 & c(\lambda) & b(\lambda) & 0\\
        d(\lambda) & 0 & 0 & a(\lambda)
    \end{pmatrix},
\end{equation}
where 
$
a(\lambda)=\rho\, \sn(\lambda + \eta ,l),\,
b(\lambda)=\rho \, \sn(\lambda ,l),\,
c(\lambda)=\rho \, \sn(\eta,l),\,
d(\lambda )=-\rho l\,\sn(\eta,l)
\,\sn(\lambda,l)
\,\sn(\lambda+\eta,l)
$
with $\rho$,$\lambda$, and $q$ as the normalization constant, rapidity, and the Hamiltonian parameter, respectively.
The local XYZ Hamiltonian term is given by,
\begin{equation}
    h_{n,n+1}=P_{n,n+1} \left.\frac{dR_{n,n+1}(\lambda)}{d\lambda}\right|_{\lambda=0}
    = \left[
    J_x \sigma_n^x \sigma_{n+1}^x
    +J_y \sigma_n^y \sigma_{n+1}^y
    +J_z \left(
    \sigma_n^z \sigma_{n+1}^z
    +
    \mathbb{I}_{n,n+1}
    \right)
    \right],
\end{equation}
where, $J_x=\rho\,(1-l\,\sn^2(\eta,l))/2,\, J_y=\rho\,(1+l\,\sn^2(\eta,l))/2,\, J_z=\rho\,\cn(\eta,l)\,\dn(\eta,l)/2$.
Furthermore, application of Landen transformation\,\cite{LandenTransformation},
\begin{equation}
    \kappa=\frac{2\sqrt{l}}{1+l},\,
    q=(1+l)\eta,
\end{equation}
transforms the Hamiltonian coefficients as follows,
\begin{equation}
    \frac{J_x}{J_y}=\,\dn(q,\kappa),\,
    \frac{J_z}{J_y}=\,\cn(q,\kappa).
\end{equation}
Thus, the Hamiltonian after Landen transformation becomes,
\begin{equation}
    H
    =
    J \sum_n \left[\,
    \dn(q,\kappa)\opS{x}{n} \opS{x}{n+1}
    + \opS{y}{n} \opS{y}{n+1}
    +\,
    \cn(q,\kappa) \opS{z}{n} \opS{z}{n+1}
    \right],
\end{equation}
where, $J=2\rho\left[1+l\,\sn^2(\eta,l)\right]$.
GZ scar states, as in Eq.\,\ref{Eq::GZScar}, are basically the eigenstates of the above Hamiltonian.

Due to the structural similarities between the Bethe algebras of the XXZ and XYZ models, we propose constructing a spectrum-generating algebra for the GZ scars in the XYZ model, analogous to the helical scars in the XXZ model, by utilizing Bethe-phantom roots.
Consequently, one must search for Bethe-phantom-root-like solutions of the Bethe ansatz equations for the XYZ model\,\cite{BetheXYZ}.

\section{\label{Appendix::Sec::More2DLattices}Lattice-Dependent and Lattice-Independent Scars in Various nearest-neighbor 2D Lattices}

\begin{figure}[t]
\includegraphics[width=\textwidth]{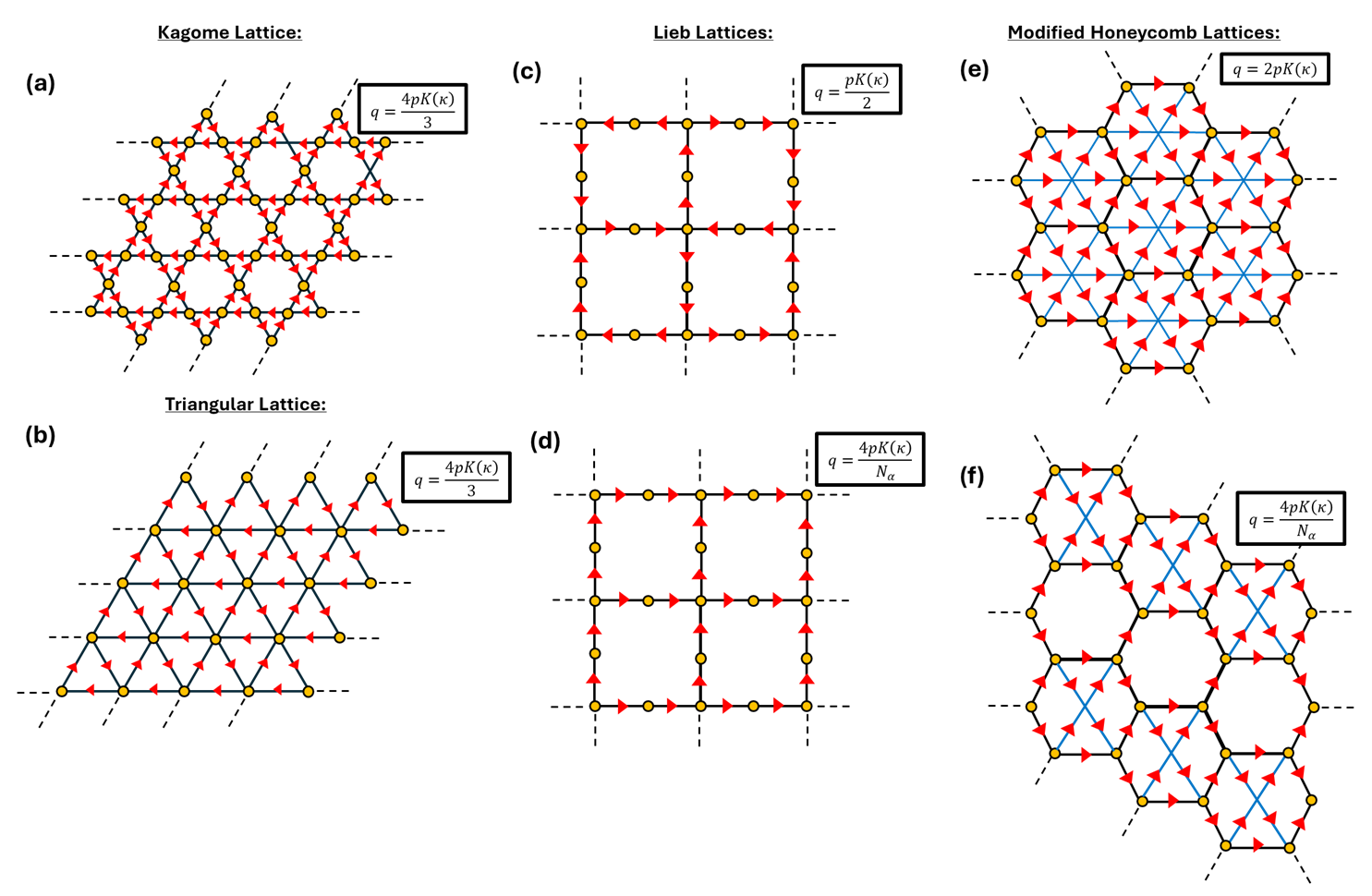}
\caption{
(a) and (b) represents lattice-dependent scar states on the kagome and triangular lattice, respectively. 
(c) and (d) illustrate lattice-dependent and lattice-independent scar states on the Lieb lattice, respectively.
(e) and (f) depict the lattice-dependent and lattice-independent scar states on modified honeycomb lattices, respectively.
}
\label{Appendix::fig::2DLattices}
\end{figure}

\subsection{Uniform Lattices}

This section demonstrates the construction of scar state on various lattices.
The only possible scar states for the uniform triangular and kagome lattices are lattice-dependent due to the presence of triangular plaquettes with an odd number of edges~(see Fig.\,\ref{Appendix::fig::2DLattices}(a), (b)).
On the other hand, the Lieb lattice can host both lattice-dependent and lattice-independent scar states, as shown in Fig.\,\ref{Appendix::fig::2DLattices}(c) and (d).
Although the uniform honeycomb lattice, in principle, cannot host any scar states due to its odd coordination number (three), it can be modified to accommodate scar states by adding additional nearest-neighbor bonds to achieve an even coordination number.
As we show in Fig.\,\ref{Appendix::fig::2DLattices}(e) and (f), the modified honeycomb lattice can host both lattice-dependent and lattice-independent scar states.

\subsection{Irregular Lattices with trimers}

\begin{figure}[t]
\includegraphics[width=\textwidth]{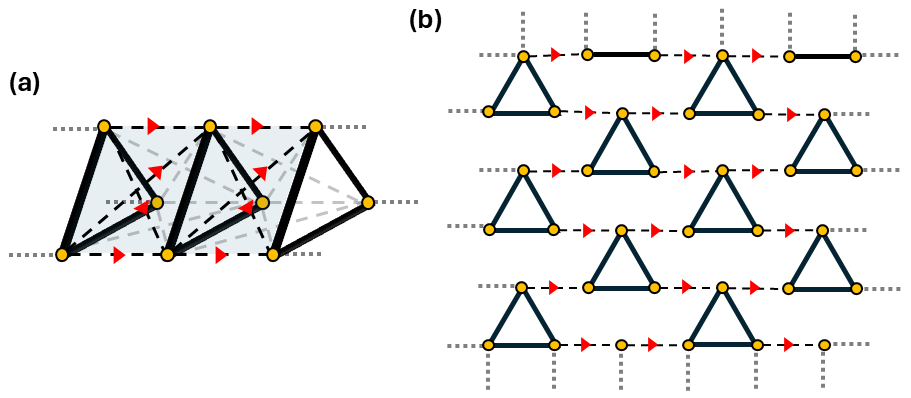}
\caption{
(a) Trimer spin-ladder model. (b) Trimer brick wall lattice model. 
The spins connected by thick black bonds interact via an isotropic Heisenberg exchange interaction, while those connected by dashed black bonds interact via an XYZ-like spin-exchange interaction.
Red arrowheads indicate a change in the $q$-value.
}
\label{Appendix::fig::FrustratedTrimerModels}
\end{figure}

In the main text, we show that incorporating $SU(2)$-invariant bonds into lattices such as triangular, kagome, and honeycomb enables the introduction of lattice-independent GZ scars in these systems.
In this appendix, we demonstrate that the concept can be further generalized to the frustrated trimer lattices.
Fig.\,\ref{Appendix::fig::FrustratedTrimerModels} illustrates two different types of trimer lattices, where thick black bonds ($\mathcal{B}_D$) form the trimers, and dashed bonds ($\mathcal{B}_d$) connect them.
The Hamiltonian of these trimer systems is,
\begin{equation}
    \hat{H}=
    J \sum_{(m,n)\in\mathcal{B}_D} \boldsymbol{\hat{S}}_m \cdot \boldsymbol{\hat{S}}_n
    +
    J^\prime \sum_{(m,n)\in\mathcal{B}_d}
    \left[
    \dn(q,\kappa)
    \opS{x}{m}\opS{x}{n}
    +\opS{y}{m}\opS{y}{n}
    +\cn(q,\kappa)\opS{z}{m}\opS{z}{n}
    \right].
\end{equation}
Red arrowheads in the Fig.\,\ref{Appendix::fig::FrustratedTrimerModels} visually represent the GZ scar.
Interestingly, no arrowheads appear on the trimer bonds $\mathcal{B}_D$, since their $q$-value remains constant.
The vertex and circuit rules continue to hold.

\clearpage
\twocolumngrid

\end{document}